\documentclass[
    aps,
    pra,
    reprint,
    nofootinbib,
    longbibliography
]{revtex4-2}

\usepackage[T1]{fontenc}
\usepackage[utf8]{inputenc}
\usepackage{lmodern}
\usepackage{microtype}
\usepackage{float}

\usepackage{amsmath}
\usepackage{amssymb}
\usepackage{mathtools}
\usepackage{braket}
\usepackage{bm}

\usepackage{graphicx}
\usepackage[caption=false]{subfig}
\usepackage{booktabs}
\usepackage[inline]{enumitem}
\usepackage{xcolor}
\usepackage{tikz}
\usetikzlibrary{arrows.meta,positioning,calc,backgrounds,fit}
\usepackage{pgfplots}
\usepgfplotslibrary{fillbetween,patchplots}
\pgfplotsset{compat=1.18}

\usepackage[bottom,hang,flushmargin]{footmisc}
\usepackage{xspace}
\usepackage{hyperref}
\IfFileExists{cleveref.sty}{
    \usepackage[capitalise,noabbrev]{cleveref}
}{}
\makeatletter
\@ifpackageloaded{cleveref}{}{%
    \newcommand{\plaq@refs}[1]{%
        \def\plaq@sep{}%
        \@for\plaq@label:=#1\do{%
            \plaq@sep\ref{\plaq@label}%
            \def\plaq@sep{, }%
        }%
    }%
    \newcommand{\cref}[1]{\plaq@refs{#1}}%
    \newcommand{\Cref}[1]{\plaq@refs{#1}}%
}
\makeatother

\definecolor{linkblue}{RGB}{24,86,145}
\definecolor{citegreen}{RGB}{32,112,83}
\definecolor{urlpurple}{RGB}{101,63,141}
\definecolor{panelgray}{RGB}{238,240,243}
\definecolor{panelblue}{RGB}{214,228,243}
\definecolor{panelgreen}{RGB}{214,238,225}
\definecolor{ncgreen}{HTML}{B2F2BB}
\definecolor{ncblue}{HTML}{A5D8FF}
\definecolor{ncred}{HTML}{FFC9C9}
\definecolor{ncyellow}{HTML}{FFEC99}
\definecolor{ncredline}{HTML}{E03131}
\definecolor{xpgreen}{HTML}{D6EFDC}
\definecolor{xpred}{HTML}{F8DAD9}
\definecolor{xpblue}{HTML}{D7E6F6}

\definecolor{xpauli}{HTML}{4CAF50}
\definecolor{stim}{HTML}{E24A44}
\definecolor{fullstate}{HTML}{A6A6A6}
\definecolor{nearclifford}{HTML}{2F80ED}
\definecolor{eqboxborder}{HTML}{E8890C} 
\definecolor{heatlow}{HTML}{F7FBFF}
\definecolor{heathigh}{HTML}{08519C}
\definecolor{iondata}{HTML}{D8DEE6}
\definecolor{ionx}{HTML}{5FD076}
\definecolor{ionz}{HTML}{FF7D7D}
\definecolor{ionpanel}{HTML}{F8FAFC}
\definecolor{ionline}{HTML}{23272F}

\hypersetup{
    colorlinks = true,
    linkcolor = linkblue,
    citecolor = citegreen,
    urlcolor = urlpurple,
    pdfauthor = {Raul Conchello Vendrell, Carlos Díaz López, Ish Dhand, Kshitij Kapoor, Davide Laureti, Marcello Massaro, Pranjal Nayak, Ivan Ogloblin, Martin B. Plenio, Shreya Prasanna Kumar, Matteo Santandrea, Varun Seshadri, Antal Száva, Trevor Vincent, Raphael Weber},
    pdftitle = {Plaquette: A hardware-aware design platform for fault-tolerant quantum computers}
}

\newcommand{\plaq}{\textsc{Plaquette}}
\newcommand{\py}[1]{\texttt{#1}}
\newcommand{\Code}{\py{Code}\xspace}
\newcommand{\Circuit}{\py{Circuit}\xspace}

\newcommand{\ErrorModel}{\py{ErrorModel}\xspace}

\newcommand{\thr}{\mathrm{th}}
\newcommand{\comp}{\mathrm{comp}}

\newcommand{\channel}{\mathcal{E}}

\newcommand{\identity}{\mathbb I}
\newcommand{\ketbra}[2]{\ket{#1}\!\bra{#2}}

\DeclareMathOperator{\Tr}{Tr}

\begin{document}

\title{Plaquette: A hardware-aware design platform for fault-tolerant quantum computers}
\author{Raul Conchello Vendrell}
\affiliation{QC Design GmbH, Lise-Meitner-Strasse 9, 89081 Ulm, Germany}
\author{Carlos Díaz López}
\affiliation{QC Design GmbH, Lise-Meitner-Strasse 9, 89081 Ulm, Germany}
\author{Ish Dhand}
\email{ish@qc.design}
\affiliation{QC Design GmbH, Lise-Meitner-Strasse 9, 89081 Ulm, Germany}
\author{Kshitij Kapoor}
\affiliation{QC Design GmbH, Lise-Meitner-Strasse 9, 89081 Ulm, Germany}
\author{Davide Laureti}
\affiliation{QC Design GmbH, Lise-Meitner-Strasse 9, 89081 Ulm, Germany}
\author{Marcello Massaro}
\affiliation{QC Design GmbH, Lise-Meitner-Strasse 9, 89081 Ulm, Germany}
\author{Pranjal Nayak}
\affiliation{QC Design GmbH, Lise-Meitner-Strasse 9, 89081 Ulm, Germany}
\author{Ivan Ogloblin}
\affiliation{QC Design GmbH, Lise-Meitner-Strasse 9, 89081 Ulm, Germany}
\author{Martin B. Plenio}
\affiliation{QC Design GmbH, Lise-Meitner-Strasse 9, 89081 Ulm, Germany}
\author{Shreya Prasanna Kumar}
\affiliation{QC Design GmbH, Lise-Meitner-Strasse 9, 89081 Ulm, Germany}
\author{Matteo Santandrea}
\affiliation{QC Design GmbH, Lise-Meitner-Strasse 9, 89081 Ulm, Germany}
\author{Varun Seshadri}
\affiliation{QC Design GmbH, Lise-Meitner-Strasse 9, 89081 Ulm, Germany}
\author{Antal Száva}
\affiliation{QC Design GmbH, Lise-Meitner-Strasse 9, 89081 Ulm, Germany}
\author{Trevor Vincent}
\affiliation{QC Design GmbH, Lise-Meitner-Strasse 9, 89081 Ulm, Germany}
\author{Raphael Weber}
\affiliation{QC Design GmbH, Lise-Meitner-Strasse 9, 89081 Ulm, Germany}
\date{\today}

\begin{abstract}
Hardware teams building fault-tolerant quantum computers (FTQCs) must decide which imperfections to suppress, and that decision requires the logical performance of the architecture under the device's actual noise.
Hardware noise often departs from the stochastic Pauli models used by scalable stabilizer simulators: superconducting transmons leak out of the computational subspace, neutral atoms scatter through intermediate states, trapped ions heat as their motional modes absorb phonons, and miscalibrated controls over-rotate coherently.
We present \plaq{}, a theoretical framework and software suite that computes the logical performance of fault-tolerant architectures directly from the physics of such imperfections.
In \plaq{}, a hardware error model is specified once, as Kraus operators, Hamiltonian--Lindblad dynamics, or an experimentally reconstructed quantum channel, and is compiled automatically into the exact or approximate representation required by each of four sampler classes: stabilizer sampling for Pauli noise, the new XPauli sampler for leakage and environment sectors, near-Clifford samplers for coherent errors, and full-state simulation for exact reference calculations.
We validate the XPauli and near-Clifford samplers against full-state simulation, which they can match within statistical uncertainty while Pauli twirling can fall short depending on the error model.
We demonstrate the framework on three error models: leakage in superconducting qubits, intermediate-state scattering in neutral atoms, and heating in trapped ions.
The size of the discrepancy between \plaq{} and Clifford-only simulations varies with platform and noise process, so reliable thresholds, error budgets, and overhead estimates require the most accurate simulation available.
\plaq{} provides a direct path from the open-system physics of a device to the logical performance of the FTQC built on it.
\end{abstract}

\maketitle

\section{Introduction}
\label{sec:introduction}

The threshold theorem tells us that imperfect hardware is not a fundamental barrier to fault-tolerant quantum computing (FTQC)~\cite{aharonov2008faulttolerant,knill1998resilient}.
If the noise on every operation is low enough, encoding the computation in an error-correcting code can suppress logical errors to any desired level.
Using the theorem to inform the design of quantum hardware, however, is not straightforward.
The theorem is stated in terms of an abstract fault rate per circuit location.
It also relies on assumptions, such as rapidly decaying probability for correlated errors, that may or may not be satisfied in real architectures.
Neither the abstract fault rate nor these assumptions refer directly to the quantities that describe a physical device.

\begin{figure}
\centering
\includegraphics[width=\linewidth]{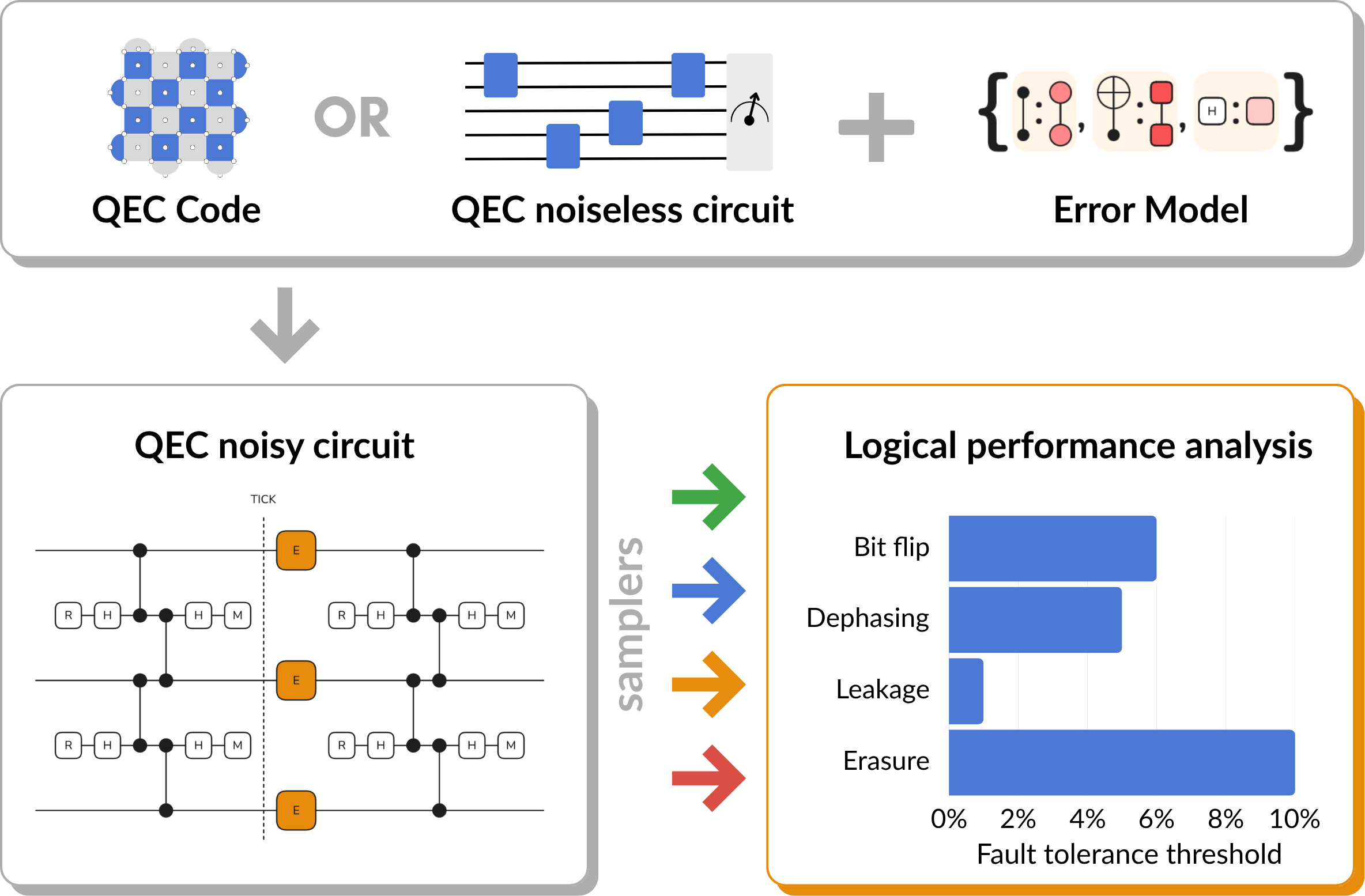}
\caption[Plaquette's workflow at a glance]{
\plaq{} at a glance.
\textbf{A QEC code} defines a \textbf{noiseless circuit} for extracting syndrome information; an \textbf{error model}, specified once as CPTP channels, is compiled into a \textbf{noisy circuit} in the exact or approximate representation that the chosen sampler requires.
This noisy circuit is sampled and decoded; the \textbf{logical performance} is returned as logical error rates, which are then aggregated into thresholds, sensitivities for different error sources, and physical-logical overheads.
}
\label{fig:glance}
\end{figure}

A hardware team building a scalable FTQC is less concerned with abstract fault rates than with concrete device parameters: coherence times, leakage rates, heating rates, scattering probabilities, detunings, gate durations, reset fidelities, and crosstalk strengths.
What the team needs to know is how each of these parameters affects the performance of the FTQC built from the device.
Engineering effort is scarce, so it should be spent on the imperfections that most limit that performance.

As a result, every hardware team faces the following practical questions:
(i) for a given device, is the noise below threshold, and if so, by how much?
(ii) which physical imperfections or sources of noise most limit the performance of the FTQC? In other words, what is the error budget~\cite{google2023suppressing} contribution of the imperfection?
(iii) given a device implementing a code and subject to an error model, what is the logical error rate?
(iv) how many resources are required per logical qubit to achieve a target residual logical error rate?

Short of building an actual FTQC, answering these questions requires simulating the encoded circuit under the device's actual noise, decoding the resulting syndromes, and extracting the logical error rate as the code is scaled.

The challenge here is that the noise of real hardware devices is not the stochastic Pauli noise assumed by the fast stabilizer simulators~\cite{gottesman_heisenberg_1998,aaronson_improved_2004,gidney2021stim} that dominate large-scale FTQC studies.
For example, a superconducting transmon qubit driven near its anharmonicity leaks into its third level~\cite{motzoi2009simple,McEwen2021}; a neutral-atom Rydberg gate scatters through an intermediate electronic state~\cite{evered2023high,saffman2010quantum}; a trapped ion or a neutral atom heats as its motional mode absorbs phonons~\cite{brownnutt2015ion,thompson2013coherence}; a spin qubit in silicon leaks into nearby valley states~\cite{yang2013spin,buterakos2021spin}; and the control fields implementing a quantum gate may be systematically miscalibrated and thus over- or under-rotate the desired unitary~\cite{bravyi2018correcting,wallman2016noise}.
None of these processes is naturally described as only a stochastic Pauli channel acting on the computational space of a qubit.

A simulator restricted to Clifford circuits and Pauli noise can map a threshold against one or two abstracted error rates, but the abstraction discards much of the physics on which the answer depends in the real world.
The existing approaches to bring realistic hardware errors into this setting can involve one or more of these techniques applied in tandem: \begin{enumerate*}[label=(\roman*), itemjoin={{; }}, itemjoin*={{; or }}]
    \item Pauli twirling of the noise model~\cite{geller2013efficient}, which replaces a general noise channel by the stochastic Pauli channel obtained by averaging over random Pauli conjugations~\cite{hann2025hybrid,PhysRevX.11.041058}
    \item performing time-dependent simulations of error channels and replacing them with depolarizing gates of similar infidelity~\cite{chadwick2025short, tham2026breakeven}
    \item attaching additional classical bits that track time correlations in noise~\cite{Regent2023highperformance,Hopfmueller2024bosonicpauli}
    \item designing simplified models that can represent features of certain error models such as time-correlations of leakage errors~\cite{Fowler2013,Suchara2014}
\end{enumerate*}.
Such reductions can be accurate, but each one demands substantial expert effort and targets a specific device and noise process.
When the noise model changes, the reduction must be rederived; validation still requires a simulation of the original channel.

Whichever of these techniques is applied, the Clifford-only simulation certifies the resulting abstraction of the device, not the device itself.
The practical design questions, i.e., whether a particular machine will reach fault tolerance, and at what physical-to-logical overhead, are then answered only as faithfully as that abstraction represents the hardware.

Closing this gap requires a design framework that accepts real hardware noise, not only Pauli noise.
We built a design framework that does just that, and implemented it in \plaq{}, a software suite for designing fault-tolerant quantum computers.
\plaq{} takes two inputs: a description of an FTQC architecture, comprising codes or circuits along with decoders, and a description of the hardware errors, given as quantum channels, that is, completely positive, trace-preserving (CPTP) maps, together with the circuit locations where they act.
From these inputs, \plaq{} returns the logical error rates, thresholds and threshold surfaces, error budgets, and overheads of that architecture under that noise.
Because the imperfections are described as channels, \plaq{} can simulate fault-tolerant architectures of hundreds to tens of thousands of qubits under realistic hardware noise, rather than under Pauli noise alone.
\Cref{fig:glance} summarizes this workflow: device physics is specified once, as a channel, and the framework evaluates the resulting logical performance, allowing rapid, accurate, and systematic exploration of design choices that lead to feasible fault-tolerance architectures.

\plaq{} is hardware agnostic, spanning the leading hardware platforms---superconducting circuits, neutral atoms, trapped ions, quantum dots, and photonics---and the three computational paradigms used to build fault-tolerant machines on them: circuit-based (CBQC), measurement-based (MBQC), and fusion-based (FBQC) quantum computing.
The present paper develops its treatment of matter-qubit platforms in the circuit model, where logical information is encoded in a code and its stabilizers are measured over many rounds; a companion paper treats photonic and spin-optical architectures, where computation proceeds by measurement and fusion~\cite{qcdesign_photonics_inprep}.

The primary contributions of this paper are threefold.
{\begin{enumerate}
\item \emph{A channel-first design framework.}
We introduce a framework for simulating and designing FTQCs under a broad class of hardware imperfections: a physical error model is specified once and is converted into the representation each sampler requires and, independently, into the stochastic detector error model that initializes the decoders included in \plaq{}.
This is the framework overviewed in \cref{fig:glance}.
\item \emph{The XPauli sampler.}
We introduce a new simulator that enables fast and efficient simulations of qubits that can leak into higher energy levels or interact with ancillary quantum systems attached to them.
\item \emph{Evidence that more accurate simulations are important for FTQC design.}
We estimate logical error rates for the same error models using several different sampling methods and show that Clifford-only simulations can provide an overly optimistic evaluation of an FTQC architecture by returning different logical error rates and thresholds as compared to those from more accurate analyses.
\end{enumerate}
\Cref{tab:features} lists the capabilities of \plaq{} at a glance.}

\begin{table}[t]
\centering
\caption[Current \plaq{} capabilities at a glance]{Current \plaq{} capabilities at a glance, with more added regularly.
}
\label{tab:features}
\newcommand{\featurecell}[1]{\parbox[t]{0.74\linewidth}{\raggedright #1}}
\begin{tabular}{@{}ll@{}}
\toprule
Codes & \featurecell{Surface, repetition, color, Shor, Steane, five-qubit, subsystem, and qLDPC families; custom codes from stabilizer data; foliation for MBQC, FBQC, \& FloBQC} \\
\addlinespace[0.45ex]
Error models & \featurecell{CPTP channels entered as Kraus operators, propagators from Hamiltonian--Lindblad dynamics, process matrices, Pauli transfer matrices, generalized-Pauli approximations, or quasiprobability decompositions; parametric, tag-dependent errors}\\
\addlinespace[0.45ex]
Samplers & \featurecell{Stim for Pauli and erasure noise; XPauli for leakage and sectors; near-Clifford for coherent and non-unital channels; full-state on CPU and GPU} \\
\addlinespace[0.45ex]
Decoders & \featurecell{PyMatching, Fusion-Blossom, Chromobius, Concat-MWPM, BP-OSD, BP-LSD, Relay-BP, belief matching, and Tesseract, initialized from detector error models; loss-aware decoding}\\
\addlinespace[0.45ex]
Analysis & \featurecell{Logical error rates, thresholds, threshold surfaces, error budgets, below-threshold suppression factors, and overhead estimates}\\
\addlinespace[0.45ex]
Platforms & \featurecell{Superconducting circuits, neutral atoms, trapped ions, quantum dots, and photonics; circuit-based, measurement-based, and fusion-based paradigms}\\
\bottomrule
\end{tabular}
\end{table}

This paper is organized as follows.
\Cref{sec:hwerrors} establishes the quantum-channel formalism used throughout the paper, derives error operations from the open-system physics of devices, and defines thresholds and threshold surfaces directly in the physical parameters that enter these error operations.
\Cref{sec:matter} describes \plaq{}'s framework that enables obtaining logical error rates of FTQC architectures under imperfections defined as arbitrary quantum channels.
\Cref{sec:usecases} showcases \plaq{} in action, showing that, under realistic non-Clifford noise, efficient error modelling that goes beyond the stabilizer formalism changes the predicted logical error rate by more than one order of magnitude, in line with full-state-vector propagation simulations; we also report threshold studies of leakage imperfections in superconducting transmon qubits and the modelling of heating dynamics in trapped-ion architectures.
The final \Cref{sec:discussion} summarizes the framework and discusses its tradeoffs, limitations, and outlook.
Together, these sections show how the physics of device imperfections, modelled as quantum channels, provides insights into the logical performance of the FTQC built from the device, expressed in physical parameters a hardware team can measure and improve.

\section{Background and framework: from device physics to logical performance}
\label{sec:hwerrors}

In \cref{sec:introduction} we framed the design problem in terms of physical quantities that hardware teams can measure or improve, such as rates, durations, detunings, couplings, temperatures, heating rates, and loss probabilities.
We collect these quantities in a vector $\bm{\eta}$.
In this paper, a device model is a parameterized prescription that assigns, for each value of $\bm{\eta}$, a noisy implementation to every circuit location in the quantum error-correction experiment.
It specifies the retained state space, the ideal target operations, and either the quantum channels themselves or the Hamiltonians, Lindblad operators, controls, durations, and rates from which those channels are computed.
For a fixed quantum error-correction experiment $\chi$, the quantity we ultimately need is the logical failure probability $P_{\mathrm{fail}}(\chi;\bm{\eta})$: the probability that decoding incorrectly predicts the parity of the logical operator when the hardware operates at parameter values $\bm{\eta}$.

This section establishes the formalism needed to go from such a device model to the target quantity $P_{\mathrm{fail}}(\chi;\bm{\eta})$.
We first describe how open-system dynamics enable calculating circuit-level quantum channels, which can be inserted into noiseless quantum circuits and simulated.
We then define logical failure for a decoded quantum error correction (QEC) experiment with these noisy circuits and explain how scaling the error-correction code provides FTQC thresholds.
Finally, we extend the same language to threshold surfaces when several hardware imperfections vary at once.
The purpose of this section is to set up the common language used by \plaq{} in \Cref{sec:matter}, where the \plaq{} formalism and sampler-specific representations are described.

\subsection{Physical dynamics as circuit-level channels}
\label{sec:channels}

The retained state space is the part of the device model on which the circuit-level channels act.
It may include only the computational qubit levels, or it may also include leakage levels, ancillary modes, or classical sectors representing an environment.
Let $\rho$ denote the density operator on this modelled state space.
For each gate, idle, reset, or measurement location, the device model then assigns a parameter-dependent physical process on this state space.
A physical process at parameters $\bm{\eta}$ is represented by a quantum channel $\channel_{\bm{\eta}}$, a linear, completely positive, trace-preserving (CPTP) map from input states to output states.
Complete positivity ensures that the map remains physical when the system is entangled with other degrees of freedom, and trace preservation conserves total probability.
For a measurement, the recorded classical outcome is included in the output register; the combined measurement map is then CPTP, although the map associated with one fixed outcome is generally trace-non-increasing.

Every CPTP channel admits a Kraus representation~\cite{nielsen_quantum_2010}
\begin{equation}
    \channel_{\bm{\eta}}(\rho)
    = \sum_j K_j(\bm{\eta})\rho K_j^{\dagger}(\bm{\eta}),
    \qquad
    \sum_j K_j^{\dagger}(\bm{\eta})K_j(\bm{\eta})=\identity,
    \label{eq:kraus}
\end{equation}
where the Kraus operators $K_j$ encode the physical process and the second relation enforces trace preservation.

A QEC experiment is implemented as a circuit, so these channels must be attached to circuit locations.
Given an ideal operation $\mathcal G$, \plaq{} represents an imperfect implementation by composing $\mathcal G$ with an error channel before or after the ideal instruction,
\begin{equation}
    \channel_{\bm{\eta}}\circ\mathcal G
    \quad\text{or}\quad
    \mathcal G\circ\channel_{\bm{\eta}},
    \label{eq:erroroperation}
\end{equation}
where the rightmost map acts first and the error channel may be the identity.
We call the inserted channel the error operation.
It reduces to the identity for a perfect implementation and otherwise describes the deviation from the target operation.
The convention allows errors to be placed before or after an ideal instruction without claiming that the underlying noise occurs instantaneously.
When noise acts continuously during a gate, the physical dynamics are integrated over the gate interval and exposed to the circuit through a channel of the form in \cref{eq:erroroperation}.
For a unitary target operation on the modelled Hilbert space, an equivalent after-gate error representation can always be obtained by composing the noisy implementation with the inverse ideal operation.

Often the channel is not supplied directly.
Instead, it is derived from a hardware-level dynamical model.
For coherent evolution with Markovian, or memoryless, dissipation, the density operator obeys the Lindblad master equation
\begin{equation}
    \begin{aligned}
        \frac{d\rho}{dt}
        &= \mathcal L[\rho] \\
        &= -i[H,\rho]
           + \sum_k L_k\rho L_k^{\dagger} \\
        &\quad - \frac{1}{2}\sum_k
           \{L_k^{\dagger}L_k,\rho\}.
    \end{aligned}
    \label{eq:lindblad}
\end{equation}
Here $H$ generates the coherent dynamics; the jump operators $L_k$ describe dissipative processes such as dephasing, heating, cooling, decay, or scattering; $[A,B]=AB-BA$ is the commutator; and $\{A,B\}=AB+BA$ is the anticommutator.
The operators and rates in \cref{eq:lindblad} depend on the physical parameters $\bm{\eta}$.
For time-independent parameters, evolution for a duration $\tau$ produces the channel generated by $e^{\mathcal L\tau}$.
For time-dependent controls, the corresponding channel is obtained by numerical integration.
This construction keeps the physical parameters explicit, so that the logical performance computed later can still be expressed in quantities a hardware team can measure or improve.
Needless to say, if a channel is reconstructed experimentally or supplied directly, this integration step is bypassed.

The CPTP channel obtained in this way contains all information retained by the chosen device model.
A full-state simulation can propagate this channel directly, but scalable QEC studies usually require approximation.
The key question is therefore which information about the channel is kept and which information is discarded.
Stabilizer simulations form the standard scalable baseline: they restrict the circuit to Clifford operations and stochastic Pauli noise.
Because Clifford operations map Pauli operators to Pauli operators under conjugation, each shot can be sampled efficiently, with runtime polynomial in the number of qubits and circuit operations~\cite{gottesman_heisenberg_1998,aaronson_improved_2004,gidney2021stim}.

This restriction leaves two distinctions that are relevant throughout the paper.
The first is whether the state is assumed to remain in the computational subspace, the two-dimensional space spanned by $\ket{0}$ and $\ket{1}$, or whether it can leak, scatter, or otherwise move into additional levels or sectors.
The second is whether the channel within the computational subspace is stochastic Pauli or contains non-Pauli structure.
Coherent over-rotations, for example, carry phase information that can interfere across the circuit.
Amplitude damping is non-unital and drives the qubit toward a preferred state.
A realistic hardware channel may contain coherent, incoherent, leakage, and environmental components at the same time.

The Pauli transfer matrix makes precise what a stochastic Pauli approximation retains from a computational-subspace channel.
For an $n$-qubit channel, its entries are
\begin{equation}
    \begin{aligned}
        [R(\channel)]_{P,Q}
        &= \frac{1}{2^n}\Tr\!\left[P\,\channel(Q)\right], \\
        &\qquad P,Q\ \text{$n$-qubit Pauli strings},
    \end{aligned}
    \label{eq:ptm}
\end{equation}
where overall Pauli phases are omitted.
A stochastic Pauli channel is diagonal in this representation.
Pauli twirling, i.e., averaging the channel over Pauli conjugations, involves setting the off-diagonal entries to zero~\cite{katabarwa2017dynamical}.
Thus, a Pauli-twirled model keeps the diagonal Pauli action that a stabilizer simulator can sample, while discarding coherent and other non-Pauli mixing carried by the off-diagonal entries.
An ordinary qubit Pauli twirl also cannot represent leaked population, because its state space contains only the computational subspace.
The resulting Pauli probabilities can still be functions of the original rates, durations, and detunings.
Thus, the same hardware parameters are used, but the simulator propagates only the Pauli-twirled channel rather than the full hardware-derived channel.
As the examples in \Cref{sec:results_error_model} show, discarding this information can substantially change the predicted logical error rate, thresholds, and error budgets depending on the device model.

\subsection{Logical failure and code scaling}
\label{sec:thresholds}

Once the circuit-level channels have been specified, the target quantity is the logical failure probability of a decoded QEC experiment.
Let $\chi$ denote the full experimental instance: the code, the syndrome-extraction circuit, the number of rounds, the tracked logical observable, and the decoder.
For a fixed vector of physical parameters $\bm{\eta}$, one simulated shot returns a classical measurement record containing all available measurement outcomes, including syndrome information and any heralded side information such as erasure or leakage flags.
It also returns the value of the tracked logical observable, or several such values when the code encodes more than one logical qubit.
The decoder uses the record, together with its model of the noise, to predict the logical value.
A logical failure occurs when this prediction is wrong.
We write the probability of that event as $P_{\mathrm{fail}}(\chi;\bm{\eta})$.
This dependence on $\chi$ becomes important when comparing a scaled family of experiments, since threshold estimates use the curves $P_{\mathrm{fail}}(\chi_i;\bm{\eta})$ for different code instances.

Estimating $P_{\mathrm{fail}}(\chi;\bm{\eta})$ therefore follows a fixed conceptual sequence.
The device model and physical parameters determine a channel at each circuit location.
Those channels define a noisy circuit.
Sampling or propagating that noisy circuit gives a distribution over records and logical outcomes.
Decoding takes these records as input and returns the logical error rate.
Repeating the same procedure over a scaled family of code instances then gives thresholds and below-threshold scaling.

A single logical error rate describes only one code instance at one point in parameter space.
It does not by itself say whether the hardware is suitable for fault-tolerant computation.
That question is answered by scaling a code family.
Consider a family of code experiments $\{\chi_1,\chi_2,\ldots\}$, ordered by increasing protection against noise, for example by increasing code distance.
Along a one-parameter scan of the physical parameters, a point is below threshold if the logical error rate decreases as the family is scaled, and above threshold if increasing the protection no longer suppresses logical failure.
The threshold is the asymptotic boundary between these two regimes.
In finite simulations it is estimated from the crossing behavior of the curves $P_{\mathrm{fail}}(\chi_i;\bm{\eta})$.
The threshold is thus a property of the complete scaling experiment, including the code and circuit family, the decoder, and the chosen family of physical channels.

Below threshold, one also needs to estimate how quickly logical failure is suppressed as the code size is increased.
For code families indexed by distance $d$, we convert the per-shot failure probability to a per-round logical error rate and fit the empirical scaling form~\cite{google2023suppressing}
\begin{equation}
    p_{L,d}^{\mathrm{round}}(\bm{\eta})
    \propto \Lambda(\bm{\eta})^{-(d+1)/2},
    \label{eq:suppression}
\end{equation}
up to a parameter-dependent prefactor that is independent of $d$ within the fitted range.
The suppression factor $\Lambda(\bm{\eta})$ is useful for estimating physical-to-logical overheads below threshold.
For code families that are not naturally indexed by distance, we compare the curves $P_{\mathrm{fail}}(\chi;\bm{\eta})$ directly rather than imposing the form in \cref{eq:suppression}.

\subsection{Threshold surfaces for several imperfections}
\label{sec:surface}

Real hardware is characterized by several physical parameters at once.
A one-dimensional threshold is therefore only a slice through a higher-dimensional parameter space.
The correctable region is the set of parameter points for which logical failure is suppressed as the code family is scaled.
The boundary of this region is the threshold surface.
This surface expresses the tradeoff among different hardware imperfections: increasing one error source may still be tolerable if another is sufficiently small, whereas a different combination of the same individual error rates may lie above threshold.

To map such a surface, \plaq{} uses one-dimensional scans along fixed directions in parameter space.
Each direction fixes the relative contribution of the different physical imperfections, and a scalar error-strength parameter moves the hardware model away from the nominal error-free point along the corresponding ray.
Reference scales, for example single-axis thresholds, can be used to put parameters with different units or natural magnitudes on comparable numerical ranges.
This parametrization chooses scan directions only; it does not assume that the physical errors combine linearly at the logical level.

A threshold scan along each ray gives a crossing point on the threshold surface.
Collecting the crossing points over a grid of directions traces the surface at the chosen resolution.
The important point for the present section is conceptual: thresholds and threshold surfaces are defined directly in the same physical parameters $\bm{\eta}$ that enter the device model.
\Cref{sec:matter} describes how \plaq{} evaluates these quantities in practice.

\section{A framework for simulating FTQC with arbitrary CPTP imperfections}
\label{sec:matter}

\begin{figure}[t]
\centering
\begin{tikzpicture}[
  >={Stealth[length=2.1mm]},
  block/.style={
    draw,
    rounded corners=1.5pt,
    line width=0.45pt,
    minimum width=28mm,
    minimum height=8mm,
    inner xsep=4pt,
    inner ysep=3pt,
    align=center,
    font=\normalsize},
  flow/.style={->, line width=0.55pt},
  note/.style={font=\scriptsize, align=center, fill=white, inner sep=1pt}]
  \node[block] (code) at (0,0) {QEC code};
  \node[block] (clean) at (0,-1.75) {Noiseless circuit};
  \node[block] (err) at (4.4,0) {Error model};
  \node[block] (noisy) at (0,-3.50) {Noisy circuit};
  \node[block] (synd) at (0,-5.25) {Syndromes};
  \node[block] (dec) at (4.4,-5.25) {Decoder};
  \node[block] (ler) at (4.4,-7.00) {LER};

  \draw[flow] (code) -- (clean);
  \draw[flow] (clean.south) -- (noisy.north);
  \draw[flow] ($(clean.south)+(0.55,0)$) -- ($(dec.north)+(-0.55,0)$);
  \draw[flow] ($(err.south)+(-0.9,0)$) -- node[note, sloped, midway] {sampler-specific\\approximation} ($(noisy.north)+(0.9,0)$);
  \draw[flow] (err.south) -- node[note, midway] {Pauli\\twirling} (dec.north);
  \draw[flow] (noisy) -- (synd);
  \draw[flow] (synd) -- (dec);
  \draw[flow] (dec) -- (ler);
\end{tikzpicture}
\caption[Plaquette schematic workflow]{Schematic of the \plaq{} simulation workflow.
A noiseless circuit is generated from a code.
The noisy circuit is compiled from the noiseless circuit and the error model in the representation required by the chosen sampler.
In parallel, a decoder model is compiled from the noiseless circuit and a Pauli-twirled version of the error model; the decoders included in \plaq{} are initialized from the resulting detector error model.
Sampling the noisy circuit produces syndromes, which are passed to the decoder; the decoder, in turn, yields the logical failure probability.}
\label{fig:plaq_workflow}
\end{figure}
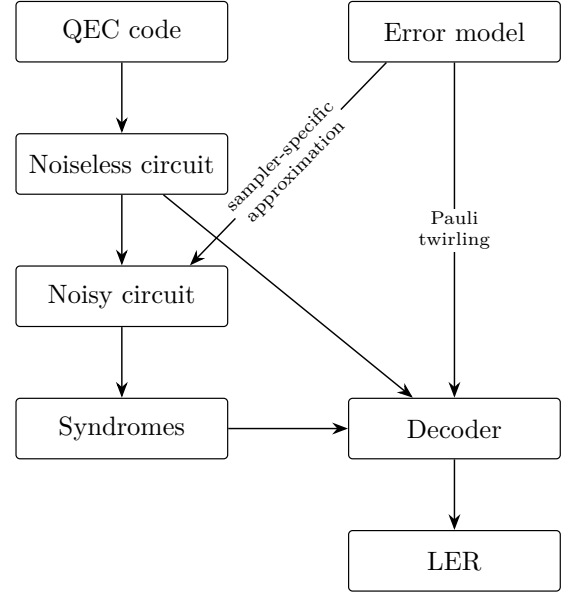

The formalism of \cref{sec:hwerrors} defines the logical failure probability $P_{\mathrm{fail}}(\chi;\bm{\eta})$ of a code experiment $\chi$ at physical parameters $\bm{\eta}$.
This section describes how \plaq{}'s framework evaluates this failure probability.
The framework does so by separating the inputs into experiment and error model: the ideal QEC experiment is specified once as a noiseless circuit; the hardware imperfections are specified once as a physical error model.
The noiseless circuit and the error model are compiled by \plaq{} into the representation required by the chosen sampler and another representation for the decoder.
The separation between the experiment and the error model allows the same physical channel to be simulated under several approximations, from stochastic Pauli sampling to full-state evolution, without rewriting the device model.

\Cref{fig:plaq_workflow} summarizes the workflow.
A code or externally supplied circuit defines the noiseless syndrome-extraction circuit.
An \ErrorModel{} assigns CPTP error operations to locations in that circuit.
Given a sampler, \plaq{} compiles these error operations into a sampler-specific noisy circuit.
The sampler generates measurement records from this noisy circuit.
A decoder is initialized from a detector error model (DEM) obtained from a Pauli-twirled noisy circuit.
Decoding predicts the tracked logical observable from those records.
Comparing the decoder prediction with the sampled logical outcome gives the logical failure probability $P_{\mathrm{fail}}(\chi;\bm{\eta})$.
Repeating this workflow over physical parameters and code instances produces the thresholds, threshold surfaces, and error budgets introduced in \cref{sec:thresholds,sec:surface}.

The rest of this section follows the workflow in the following order.
We first describe circuits, which define the ideal QEC experiment.
We then describe error models, which attach physical channels to circuit locations.
Next we describe the sampler hierarchy, which determines what information from those channels is propagated through the circuit.
We then describe decoder models and loss-aware decoding.
Finally, we describe the analysis layer that uses sampled logical error rates to obtain thresholds and resource-scaling estimates.

\subsection{Circuits}
\label{sec:circuits}

A \plaq{} \Circuit is an ordered sequence of circuit instructions.
Its instruction set is a strict superset of Stim's~\cite{gidney2021stim}: any Stim-compatible circuit runs natively in \plaq{}, while additional instructions support channels and operations that are outside the usual Clifford-Pauli setting.
These include generalized-Pauli channels for leakage and sector transitions, general unitary and Kraus channels for full-state simulations, projectors, and native non-Clifford gates such as $T$.

The key distinction is between a \emph{noiseless circuit} and a \emph{noisy circuit}.
A noiseless circuit specifies the ideal measurement schedule of the QEC experiment: the gates, resets, measurements, detectors, and logical observables that would be present in the absence of hardware imperfections.
It is the template from which all sampler-specific circuits are compiled.
A noisy circuit is the object actually sampled.
It contains the same ideal QEC structure, together with error instructions inserted according to the chosen physical error model and sampler representation.

Noiseless circuits can be imported from external text files or generated programmatically from \Code objects.
A \Code stores the stabilizers or gauge generators, the logical operators, and the metadata needed to generate a syndrome-extraction circuit, such as the measurement schedule, ancilla assignment, and two-qubit-gate ordering.
This provides a single interface for constructing memory experiments and other QEC circuits across different code families~\cite{calderbank_good_1996,steane_error_1996,bravyi2012subsystem,higgott2021subsystem,bombin2006topological,lee2025colorscaling}.
The built-in library includes standard surface, repetition, color, Shor, Steane, and five-qubit codes, together with subsystem and qLDPC families~\cite{bravyi_quantum_1998,horsman_surface_2012,bombin2006topological,lee2025colorscaling,shor_scheme_1995,steane_error_1996,laflamme_perfect_1996,bravyi2012subsystem,higgott2021subsystem,breuckmann2021quantum,kitaev_quantum_1997,tillich2013quantum,Bravyi2024HighThreshold,panteleev2021degenerate,panteleev2021quantum,Aasen2025}.
Custom codes can also be supplied from stabilizer data or CSS parity-check matrices.
The results in this paper use only a subset of this library, but the same circuit interface is used throughout.

Given a QEC code, \plaq{} can programmatically generate a syndrome-extraction circuit acting on physical qubits to perform memory or other simple Clifford logical operations.
In order to do so, it generates the circuit one `chunk' at a time, using stimflow~\cite{gidney2021stim}.
The chunk is a self-contained circuit that includes information about its `flows', which make explicit the input and output stabilizers and logicals for this chunk.
Several chunks, e.g., corresponding to different rounds of error correction, are stitched together into the full circuit; their internal correctness and composability are validated, and detectors are derived automatically.
Thus we ensure that the circuit generated is a legitimate syndrome-extraction circuit that corresponds to the intended QEC code. 

Noisy circuits may also be supplied directly.
In the typical workflow, however, the user supplies a noiseless circuit and an \ErrorModel{}, and \plaq{} constructs the noisy circuit automatically.
This keeps the architecture and the device physics independent: the same QEC circuit can be simulated under several hardware models, and the same hardware model can be compiled into several sampler representations.

\subsection{Error models}
\label{sec:errors}

An \ErrorModel{} specifies which error operation acts at each relevant location of a noiseless circuit.
Errors can be attached before or after gates, at the start of measurement rounds, by gate type, by gate tag, or by combinations of these selectors.
The object therefore describes both the physical channel and where that channel acts in the QEC experiment.

The channels in an \ErrorModel{} can be initialized from several equivalent or derived representations: Kraus operators; a propagator, often obtained by integrating the master equation \cref{eq:lindblad}; a process matrix in a Pauli basis; a Pauli transfer matrix; a generalized-Pauli approximation that tracks leaked levels and sectors; or a quasiprobability decomposition over Clifford channels.
\plaq{} converts between these representations and performs approximations as needed, typically routing through the Kraus form.
Thus, a device model may enter as Hamiltonian and Lindblad operators, as an experimentally reconstructed channel, or as a directly specified circuit-level channel.

The same interface also supports \emph{parametric errors}.
In this case, the channel parameters are functions of information stored in a gate tag.
Such a construction is valuable if the error models depend on an intermediate compilation step that takes the noiseless circuit and turns it into a compiled version that furthermore contains information about where errors are introduced and by how much.
For example, a circuit compiler may tag a shuttling operation by its duration or distance, and the corresponding error channel can read that tag to assign a location-specific heating, loss, or dephasing probability.
This keeps spatially or temporally varying hardware imperfections inside the same physical-parameter language used in \cref{sec:hwerrors}.

The decisive feature, which is central to \plaq{}'s framework, is that the physical error model is specified once, while the sampler representation is chosen later.
Given the same input channel, \plaq{} can produce a Pauli-twirled channel for the stabilizer sampler, a generalized-Pauli channel for XPauli, a quasiprobability decomposition for the near-Clifford samplers, or the full Kraus channel for full-state simulation.
The examples in \cref{sec:results_error_model} use this feature directly: one physical error model is compiled into several representations, and the resulting logical error rates are compared.

Given its flexibility, \plaq{} can handle a wide variety of relevant hardware imperfections.
These include single-qubit Pauli errors, gate-level noise, leakage to higher energy levels, and environment-dependent errors, or other CPTP channels that are relevant in FTQC design.
For benchmarking and common studies, \plaq{} also provides standard circuit-level error models, including uniform depolarizing noise and the superconducting-inspired SI1000 model~\cite{gidney2021honeycomb}.

\subsection{Samplers}
\label{sec:samplers}

\begin{table}[t]
\centering
\caption[The four Plaquette sampler classes]{The four \plaq{} sampler classes, organized by the information each retains from the original CPTP channel.
Each sampler consumes the channel representation that the error layer prepares for it (\Cref{sec:errors}) and is exposed as both a measurement and a detector sampler.
The near-Clifford class groups four methods that exploit Clifford structure while handling residual non-Clifford operations differently, and the full-state class is split by CPU and GPU backend; the indented rows list these routes individually, and \cref{sec:ncsamplers} describes each near-Clifford route in detail.}
\label{tab:samplers}
\newcommand{\samplercaptures}[1]{\parbox[t]{0.30\columnwidth}{\raggedright #1}}
\newcommand{\samplerbackend}[1]{\parbox[t]{0.39\columnwidth}{\raggedright #1}}
\begin{tabular}{@{}lll@{}}
\toprule
Sampler & Captures & Backend \\
\midrule
Stim & \samplercaptures{Pauli, Erasure} & \samplerbackend{Stim~\cite{gidney2021stim}} \\
\specialrule{0.35pt}{0.45ex}{0.45ex}
XPauli & \samplercaptures{Leakage, Scattering, Heating, Cosmic ray bursts, etc.} & \samplerbackend{Extended tableau} \\
\specialrule{0.35pt}{0.45ex}{0.45ex}
\multicolumn{3}{@{}l@{}}{Near-Clifford sampling:} \\
\quad general & \samplercaptures{Coherent and non-unital channels} & \samplerbackend{Quasiprobability sampling over Clifford channels~\cite{Bennink2017,Hakkaku2021}} \\
\quad unitary & \samplercaptures{Coherent (unitary) errors} & \samplerbackend{Operator-level Clifford decomposition in CH form~\cite{Bravyi2019,LeBlond2025}} \\
\quad \textit{tsim} & \samplercaptures{Native non-Clifford gates} & \samplerbackend{Stabilizer-rank / ZX simulation via \textit{tsim}~\cite{haenel2026tsim}} \\
\quad \textit{Clifft} & \samplercaptures{Native non-Clifford gates} & \samplerbackend{Frame-factored active-subspace simulation via \textit{Clifft}~\cite{chase2026clifft}} \\
\specialrule{0.35pt}{0.45ex}{0.45ex}
\multicolumn{3}{@{}l@{}}{{Full-state sampling:}} \\
\quad CPU & \samplercaptures{Arbitrary CPTP operations} & \samplerbackend{Cirq} \\
\quad GPU & \samplercaptures{Arbitrary CPTP operations} & \samplerbackend{cuQuantum} \\
\bottomrule
\end{tabular}
\end{table}

A sampler draws measurement records from a noisy circuit.
The sampler hierarchy in \Cref{tab:samplers} is organized by the information retained from the original CPTP channel.
The Stim backend retains stochastic Pauli and erasure information.
XPauli retains stochastic transitions among computational, leaked, and sector labels, together with Pauli errors inside the computational subspace.
The near-Clifford samplers retain selected coherent or non-Clifford structure through Clifford decompositions.
The full-state samplers propagate the complete channel, at exponential cost.

The important point is that these samplers receive as input different representations of the same physical model.
A hardware-derived channel can therefore be run through several samplers to measure how much a chosen approximation changes the predicted logical error rate.
\Cref{tab:samplers} also summarizes the representation and implementation associated with each sampler class.
The following subsections describe the three sampler classes that go beyond ordinary stabilizer sampling.
For the Stim backend, which implements tableau-based stabilizer simulation and Pauli-frame sampling for Clifford circuits with Pauli noise, we refer to~\citet{gidney2021stim}.

\subsubsection{The XPauli sampler}
\label{sec:xpauli}

\newcommand{\xplvl}[5]{%
  \draw[line width=1.1pt, black!75, line cap=round] (#2,#4) -- (#3,#4);
  \node[anchor=center, font=\small, text=black!75] at ({(#1+#2)/2},#4) {$#5$};}
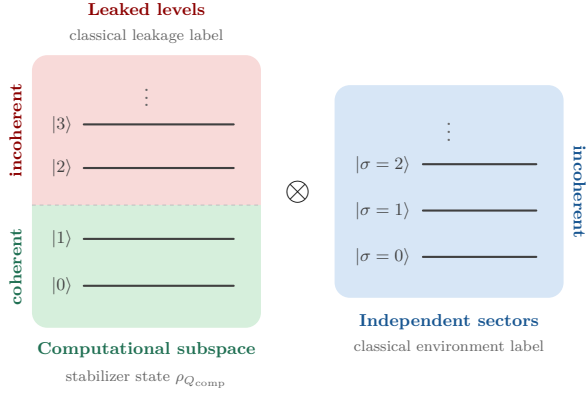
\begin{figure}[t]
\centering
\resizebox{0.9\columnwidth}{!}{%
\begin{tikzpicture}[font=\small]
  \def\Lx{-0.15}\def\Lw{4.20}       
  \def\splity{2.10}                 
  \def\Lb{-0.15}\def\Lt{4.85}       
  \def\linL{0.80}\def\linR{3.55}    
  \def\lzone{0.0}                   
  \def\Rx{5.40}\def\Rw{4.20}        
  \def\Rb{0.40}\def\Rt{4.30}        
  \def\rinL{7.02}\def\rinR{9.10}    
  \def\rzone{5.55}                  
  \def\rad{8pt}                     
  \begin{scope}
    \clip[rounded corners=\rad] (\Lx,\Lb) rectangle ({\Lx+\Lw},\Lt);
    \fill[xpgreen] (\Lx,\Lb) rectangle ({\Lx+\Lw},\splity);
    \fill[xpred]   (\Lx,\splity) rectangle ({\Lx+\Lw},\Lt);
  \end{scope}
  \draw[black!25, line width=0.5pt, dash pattern=on 2pt off 2pt] (\Lx,\splity) -- ({\Lx+\Lw},\splity);
  \xplvl{\lzone}{\linL}{\linR}{0.62}{\ket{0}}
  \xplvl{\lzone}{\linL}{\linR}{1.48}{\ket{1}}
  \xplvl{\lzone}{\linL}{\linR}{2.78}{\ket{2}}
  \xplvl{\lzone}{\linL}{\linR}{3.58}{\ket{3}}
  \node[font=\large, text=black!60] at ({\Lx+\Lw/2},4.18) {$\vdots$};
  \node[rotate=90, font=\small\bfseries, text=citegreen]   at ({\Lx-0.30},{(\Lb+\splity)/2})  {coherent};
  \node[rotate=90, font=\small\bfseries, text=red!55!black] at ({\Lx-0.30},{(\splity+\Lt)/2}) {incoherent};
  \node[font=\LARGE] at ({(\Lx+\Lw+\Rx)/2},{(\Lb+\Lt)/2}) {$\otimes$};
  \begin{scope}
    \clip[rounded corners=\rad] (\Rx,\Rb) rectangle ({\Rx+\Rw},\Rt);
    \fill[xpblue] (\Rx,\Rb) rectangle ({\Rx+\Rw},\Rt);
  \end{scope}
  \xplvl{\rzone}{\rinL}{\rinR}{1.15}{\ket{\sigma=0}}
  \xplvl{\rzone}{\rinL}{\rinR}{2.00}{\ket{\sigma=1}}
  \xplvl{\rzone}{\rinL}{\rinR}{2.85}{\ket{\sigma=2}}
  \node[font=\large, text=black!60] at ({\Rx+\Rw/2},3.55) {$\vdots$};
  \node[rotate=-90, font=\small\bfseries, text=linkblue] at ({\Rx+\Rw+0.30},{(\Rb+\Rt)/2}) {incoherent};
  \node[align=center, text=citegreen,  font=\bfseries]            at ({\Lx+\Lw/2},{\Lb-0.41}) {Computational subspace};
  \node[align=center, font=\footnotesize, text=black!60]          at ({\Lx+\Lw/2},{\Lb-0.94}) {stabilizer state $\rho_{Q_{\comp}}$};
  \node[align=center, text=red!55!black, font=\bfseries]          at ({\Lx+\Lw/2},{\Lt+0.85}) {Leaked levels};
  \node[align=center, font=\footnotesize, text=black!60]          at ({\Lx+\Lw/2},{\Lt+0.34}) {classical leakage label};
  \node[align=center, text=linkblue,   font=\bfseries]            at ({\Rx+\Rw/2},{\Rb-0.43}) {Independent sectors};
  \node[align=center, font=\footnotesize, text=black!60]          at ({\Rx+\Rw/2},{\Rb-0.87}) {classical environment label};
\end{tikzpicture}%
}
\caption[The XPauli state representation]{State representation underlying the XPauli sampler.
Each qubit carries an extended label rather than a bare two-level state.
(i)~Qubits in the computational subspace spanned by $\ket{0}$ and $\ket{1}$ (green) keep full coherence and are stored jointly as one stabilizer state; their level label reads $\comp$, short for `computational'.
(ii)~A qubit in a leaked level (red) is instead tracked by a classical level label $x_i \in \{2,3,\dots\}$.
(iii)~A separate sector label $\sigma_i$ (blue) records the classical state of the qubit's environment.
Leaked levels and sectors are tracked classically: coherence is confined to the computational subspace, which is what keeps the simulation efficient.
The explicit state representation is given in the main text.}
\label{fig:xpauli}
\end{figure}

The XPauli sampler extends stabilizer simulation to systems whose qubits can leave the computational subspace or carry additional classical environment labels.
It is designed for noise in which coherence inside the computational subspace must be retained, but coherence between different leaked levels, sectors, or environment labels can be discarded~\cite{Marshall2025incoherent}.
This assumption enables a classical-jump approach to simulations, i.e., tracking non-computational degrees of freedom via classical labels while preserving the efficient stabilizer description of the information in the computational subspace.

The sampler represents a state by the triple $(\rho_{Q_{\comp}},\bm{x},\bm{\sigma})$.
Here $\rho_{Q_{\comp}}$ is a stabilizer state on the set $Q_{\comp}$ of qubits currently in the computational subspace; $x_i\in\{\comp,2,3,\ldots\}$ records whether qubit $i$ is computational ($x_i=\comp$) or in a leaked level; and $\sigma_i\in\{0,1,\ldots\}$ records an additional sector label, such as an environmental or motional state.
The physical state represented by the sampler is therefore a stabilizer state on the computational qubits, tensored with classical labels for the remaining degrees of freedom,
\[
    \rho_{Q_{\comp}} \otimes \bigotimes_{i\notin Q_{\comp}}\ketbra{x_i}{x_i},
\]
with the sectors carried as classical metadata.
\Cref{fig:xpauli} illustrates this approximate state representation.

An XPauli channel is a transition table between these labels.
Each transition may also carry a Pauli error on the qubits that remain in the computational subspace.
For example, leakage is a transition $\comp\to2$, leak-down is a transition $2\to\comp$, and a change in an environment label is a transition such as $(0,\comp)\to(1,\comp)$.
The same table can describe correlated events, such as a Pauli error on one qubit conditioned on the leakage state of another.
\plaq{} derives these transition tables from Kraus channels by generalized Pauli twirling~\cite{google2023suppressing,Hopfmueller2024bosonicpauli}.

Sampling proceeds in two stages.
First, the sampler draws a trajectory of subspace and sector labels through the circuit.
This trajectory records when each qubit leaks, returns to the computational subspace, or changes sector.
Because these transitions depend only on the label state and the channel tables, this stage is a classical random walk over the circuit locations.
Second, the trajectory determines an effective Clifford circuit on the qubits that are computational at each location.
That Clifford circuit is sampled by stabilizer methods.

The rules for constructing the effective Clifford circuit follow from the incoherence assumption.
When a qubit leaves the computational subspace, it is removed from the stabilizer state, which decoheres it from the remaining register.
While it is leaked, gates acting on it are skipped, and two-qubit gates are skipped whenever the prescribed physical model says that a leaked participant prevents the ideal computational gate from acting.
XPauli channels may still apply residual Pauli errors to neighboring qubits, including errors conditioned on the leaked label.
When a qubit returns to the computational subspace, its computational state is set by the return rule encoded in the channel or by the chosen readout/return strategy.
A measurement of a leaked qubit is also configurable: it may be projected to a random binary outcome, reported as a raw level, or mapped deterministically to a chosen computational outcome.

The cost above ordinary stabilizer simulation is the trajectory bookkeeping and the explicit tableau evolution of the resulting Clifford circuits.
The overall cost therefore scales like that of a Clifford simulation, polynomially in the number of qubits and circuit locations, rather than exponentially as in full-state simulation.
Two Stim features that improve performance are not available in this method of sampling.
Detector-error-model sampling and Pauli-frame sampling assume a fixed detector structure that is deterministic under noiseless execution.
The effective Clifford circuits generated by an XPauli trajectory need not have that property.
XPauli simulations therefore use explicit tableau evolution of the trajectory-conditioned Clifford circuit.

Now we put XPauli into context with respect to alternative simulation methods.
Alternative Clifford-based simulation methods for leakage include those of Refs.~\cite{Fowler2013,Suchara2014}, which keep track of leaked qubits with classical labels and act as a depolarizing error on partner qubits that a leaked qubit interacts with.
Compared with Ref.~\cite{Fowler2013}, and even within the restricted leakage setting, XPauli can account for significantly more errors such as skipped gates, leakage spread, correlated Pauli-leakage errors, and biased Pauli errors.
XPauli includes as edge cases functionality from both the Pauli+ simulators~\cite{google2023suppressing} that handle leakage in superconducting qubits, and the Bosonic Pauli+ simulator that uses sectors to study concatenated finite-energy GKP simulations~\cite{Hopfmueller2024bosonicpauli}.
Other alternative approaches include tensor-network leakage simulation, which can capture coherent leakage but is more suitable for 1D codes with feasible bond dimensions~\cite{Manabe2025efficient}, and full-state leakage simulations that are infeasible to scale beyond a small retained qudit Hilbert space~\cite{camps2026leakage}.

\subsubsection{The near-Clifford samplers}
\label{sec:ncsamplers}

\newcommand{\nchl}[2]{\tikz[baseline=(nchlx.base)]{%
  \node[inner sep=2pt, rounded corners=1.5pt, fill=#1] (nchlx) {$#2$};}}
\newcommand{\ncgen}{\nchl{ncgreen}{I\cdot\rho\cdot I^{\dagger}}}
\newcommand{\ncgenB}{\nchl{ncblue}{S\cdot\rho\cdot S^{\dagger}}}
\newcommand{\ncgenR}{\nchl{ncred}{Z\cdot\rho\cdot Z^{\dagger}}}
\newcommand{\ncuni}[4]{\nchl{#1}{#2}$\,\cdot\rho\cdot\,$\nchl{#3}{#4^{\dagger}}}
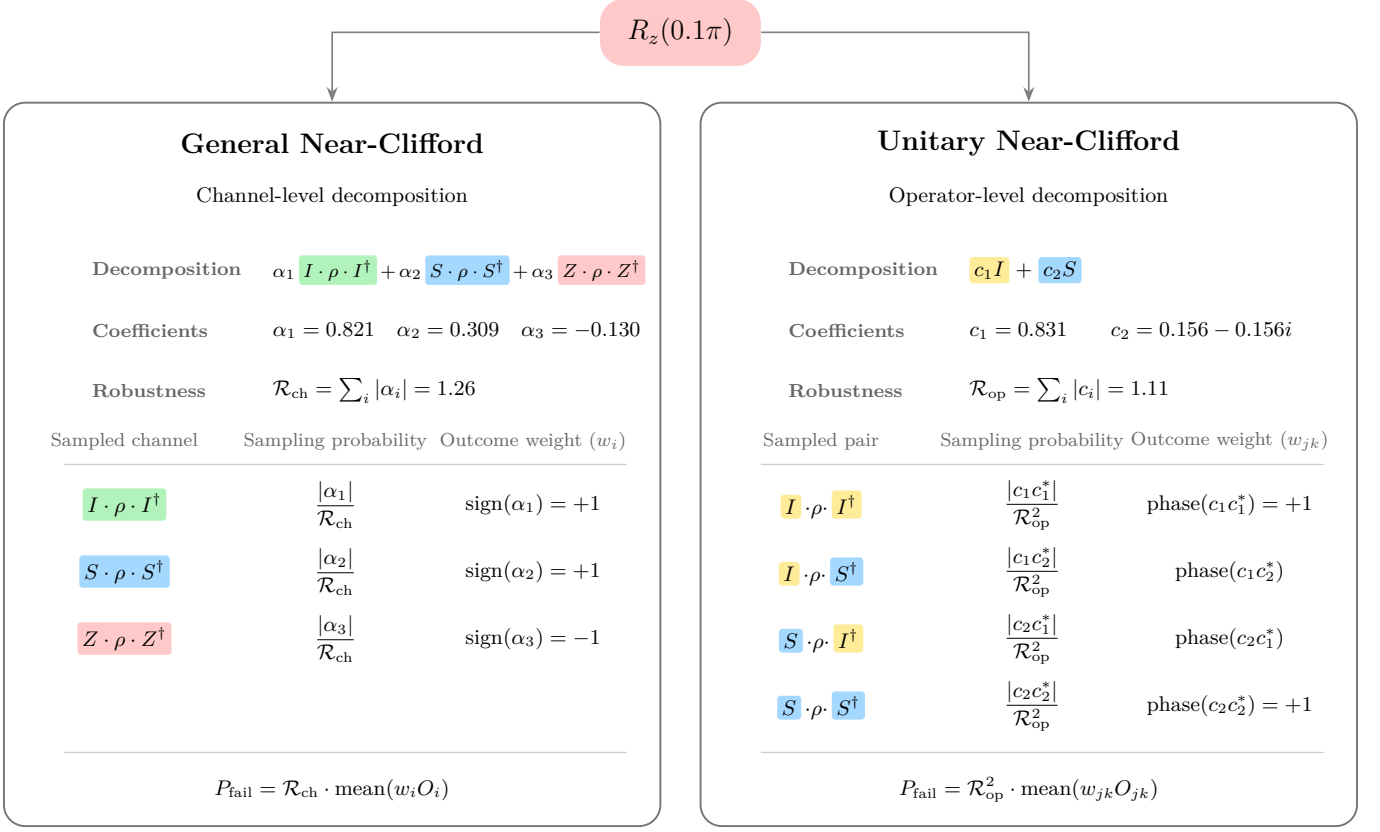
\begin{figure*}[t]
\centering
\resizebox{\textwidth}{!}{%
\begin{tikzpicture}[
  >={Stealth[length=2.2mm]},
  font=\small,
  panel/.style={draw=black!55, line width=0.8pt, rounded corners=9pt, fill=white},
  gatebox/.style={draw=none, rounded corners=10pt,
    fill=ncred, minimum width=24mm, minimum height=10mm, font=\large},
  ptitle/.style={font=\bfseries\large},
  psub/.style={font=\small},
  colhead/.style={font=\footnotesize, text=black!60},
  rowlbl/.style={font=\footnotesize\bfseries, text=black!60, anchor=west},
  rowval/.style={anchor=west},
  ler/.style={font=\small}]
  \def\cxL{-5.2}
  \def\cxR{5.2}
  \def\pw{9.8}
  \def\ph{10.8}
  \def\pcy{0.6}
  \def\dterm{-3.1}
  \def\dprob{0.05}
  \def\dwt{3.0}
  \def\dlbl{-3.7}
  \def\dval{-1.0}
  \node[panel, minimum width=\pw cm, minimum height=\ph cm] (lpanel) at (\cxL,\pcy) {};
  \node[panel, minimum width=\pw cm, minimum height=\ph cm] (rpanel) at (\cxR,\pcy) {};
  \node[gatebox] (gate) at (0,7.05) {$R_{z}(0.1\pi)$};
  \draw[->, black!50, line width=0.7pt] (gate.west) -| (lpanel.north);
  \draw[->, black!50, line width=0.7pt] (gate.east) -| (rpanel.north);

  \node[ptitle] at (\cxL,5.4) {General Near-Clifford};
  \node[psub]   at (\cxL,4.6) {Channel-level decomposition};
  \node[rowlbl] at ({\cxL+\dlbl},3.5) {Decomposition};
  \node[rowval, font=\footnotesize] at ({\cxL+\dval},3.5)
    {$\alpha_{1}\,$\ncgen$\,+\,\alpha_{2}\,$\ncgenB$\,+\,\alpha_{3}\,$\ncgenR};
  \node[rowlbl] at ({\cxL+\dlbl},2.6) {Coefficients};
  \node[rowval] at ({\cxL+\dval},2.6) {$\alpha_{1}=0.821 \quad \alpha_{2}=0.309 \quad \alpha_{3}=-0.130$};
  \node[rowlbl] at ({\cxL+\dlbl},1.7) {Robustness};
  \node[rowval] at ({\cxL+\dval},1.7) {$\mathcal R_{\mathrm{ch}}=\sum_{i}|\alpha_{i}|=1.26$};
  \node[colhead] at ({\cxL+\dterm},0.95) {Sampled channel};
  \node[colhead] at ({\cxL+\dprob},0.95) {Sampling probability};
  \node[colhead] at ({\cxL+\dwt},0.95)   {Outcome weight ($w_{i}$)};
  \draw[black!20, line width=0.5pt] ({\cxL+\dterm-0.9},0.6) -- ({\cxL+\dwt+1.4},0.6);
  \node at ({\cxL+\dterm},0.0)  {\ncgen};
  \node at ({\cxL+\dprob},0.0)  {$\dfrac{|\alpha_{1}|}{\mathcal R_{\mathrm{ch}}}$};
  \node at ({\cxL+\dwt},0.0)    {$\operatorname{sign}(\alpha_{1})=+1$};
  \node at ({\cxL+\dterm},-1.0) {\ncgenB};
  \node at ({\cxL+\dprob},-1.0) {$\dfrac{|\alpha_{2}|}{\mathcal R_{\mathrm{ch}}}$};
  \node at ({\cxL+\dwt},-1.0)   {$\operatorname{sign}(\alpha_{2})=+1$};
  \node at ({\cxL+\dterm},-2.0) {\ncgenR};
  \node at ({\cxL+\dprob},-2.0) {$\dfrac{|\alpha_{3}|}{\mathcal R_{\mathrm{ch}}}$};
  \node at ({\cxL+\dwt},-2.0)   {$\operatorname{sign}(\alpha_{3})=-1$};
  \draw[black!20, line width=0.5pt] ({\cxL+\dterm-0.9},-3.7) -- ({\cxL+\dwt+1.4},-3.7);
  \node[ler] at (\cxL,-4.25) {$P_{\mathrm{fail}}=\mathcal R_{\mathrm{ch}}\cdot\mathrm{mean}(w_{i}O_{i})$};

  \node[ptitle] at (\cxR,5.4) {Unitary Near-Clifford};
  \node[psub]   at (\cxR,4.6) {Operator-level decomposition};
  \node[rowlbl] at ({\cxR+\dlbl},3.5) {Decomposition};
  \node[rowval] at ({\cxR+\dval},3.5) {\nchl{ncyellow}{c_{1} I}$\;+\;$\nchl{ncblue}{c_{2} S}};
  \node[rowlbl] at ({\cxR+\dlbl},2.6) {Coefficients};
  \node[rowval] at ({\cxR+\dval},2.6) {$c_{1}=0.831 \qquad c_{2}=0.156-0.156i$};
  \node[rowlbl] at ({\cxR+\dlbl},1.7) {Robustness};
  \node[rowval] at ({\cxR+\dval},1.7) {$\mathcal R_{\mathrm{op}}=\sum_{i}|c_{i}|=1.11$};
  \node[colhead] at ({\cxR+\dterm},0.95) {Sampled pair};
  \node[colhead] at ({\cxR+\dprob},0.95) {Sampling probability};
  \node[colhead] at ({\cxR+\dwt},0.95)   {Outcome weight ($w_{jk}$)};
  \draw[black!20, line width=0.5pt] ({\cxR+\dterm-0.9},0.6) -- ({\cxR+\dwt+1.4},0.6);
  \node at ({\cxR+\dterm},0.0)  {\ncuni{ncyellow}{I}{ncyellow}{I}};
  \node at ({\cxR+\dprob},0.0)  {$\dfrac{\left|c_{1}c_{1}^{*}\right|}{\mathcal R_{\mathrm{op}}^{2}}$};
  \node at ({\cxR+\dwt},0.0)    {$\operatorname{phase}(c_{1}c_{1}^{*})=+1$};
  \node at ({\cxR+\dterm},-1.0) {\ncuni{ncyellow}{I}{ncblue}{S}};
  \node at ({\cxR+\dprob},-1.0) {$\dfrac{\left|c_{1}c_{2}^{*}\right|}{\mathcal R_{\mathrm{op}}^{2}}$};
  \node at ({\cxR+\dwt},-1.0)   {$\operatorname{phase}(c_{1}c_{2}^{*})$};
  \node at ({\cxR+\dterm},-2.0) {\ncuni{ncblue}{S}{ncyellow}{I}};
  \node at ({\cxR+\dprob},-2.0) {$\dfrac{\left|c_{2}c_{1}^{*}\right|}{\mathcal R_{\mathrm{op}}^{2}}$};
  \node at ({\cxR+\dwt},-2.0)   {$\operatorname{phase}(c_{2}c_{1}^{*})$};
  \node at ({\cxR+\dterm},-3.0) {\ncuni{ncblue}{S}{ncblue}{S}};
  \node at ({\cxR+\dprob},-3.0) {$\dfrac{\left|c_{2}c_{2}^{*}\right|}{\mathcal R_{\mathrm{op}}^{2}}$};
  \node at ({\cxR+\dwt},-3.0)   {$\operatorname{phase}(c_{2}c_{2}^{*})=+1$};
  \draw[black!20, line width=0.5pt] ({\cxR+\dterm-0.9},-3.7) -- ({\cxR+\dwt+1.4},-3.7);
  \node[ler] at (\cxR,-4.25) {$P_{\mathrm{fail}}=\mathcal R_{\mathrm{op}}^{2}\cdot\mathrm{mean}(w_{jk}O_{jk})$};
\end{tikzpicture}%
}
\caption[The near-Clifford sampling pipelines]{The two quasiprobability near-Clifford samplers, illustrated on the coherent rotation $R_{z}(0.1\pi)$.
(Left) The general sampler decomposes the \emph{channel} as $\channel=\sum_k\alpha_k\Phi_k$, where $\Phi_k$ denotes a Clifford channel; each shot draws one channel with probability $|\alpha_k|/\mathcal R_{\mathrm{ch}}$ and carries the sign of $\alpha_k$ as its weight.
(Right) The unitary sampler decomposes the \emph{operator} as $U=\sum_i c_i C_i$, where $C_i$ denotes a Clifford unitary; each shot draws a left/right pair $(C_j,C_k)$ with probability $|c_jc_k^*|/\mathcal R_{\mathrm{op}}^2$ and carries the phase of $c_jc_k^*$ as its weight.
The estimators multiply the weighted sample mean $\overline{wO}$ by the total robustness accumulated over all sampled locations.}
\label{fig:ncsamplers}
\end{figure*}

\plaq{}'s near-Clifford samplers address a different limitation of stabilizer simulation.
Coherent errors and native non-Clifford gates cannot be captured by the Clifford-only or Clifford plus classical jump approach of \cref{sec:xpauli}.
A single non-Clifford element can prevent direct stabilizer simulation and, in general, force full-state evolution.
The near-Clifford samplers avoid that cost by expanding the non-Clifford content over Clifford elements.
Each sampled shot is then a Clifford circuit, while the average over weighted shots reconstructs the target channel or unitary.

\plaq{} provides four near-Clifford samplers.
The first is a general channel-level quasiprobability sampler~\cite{Bennink2017,Hakkaku2021}.
The second is an operator-level sampler specialized to unitary coherent errors~\cite{LeBlond2025}.
The third wraps the external stabilizer-rank/ZX-based \textit{tsim} simulator~\cite{haenel2026tsim}.
The fourth wraps \textit{Clifft}, a frame-factored exact near-Clifford simulator whose dense state-vector evolution is confined to a dynamic active subspace~\cite{chase2026clifft}.
\Cref{fig:ncsamplers} illustrates the first two routes on a coherent rotation.

The general near-Clifford sampler decomposes a channel as a signed sum of Clifford channels,
\begin{equation}
    \label{eq:near-clifford}
    \channel = \sum_k \alpha_k\,\Phi_k,
    \qquad
    \mathcal R_{\mathrm{ch}} = \sum_k |\alpha_k| \ge 1,
\end{equation}
where $\Phi_k$ is a Clifford channel and the real coefficients $\alpha_k$ may be negative.
When converting an error model to this representation, \plaq{} finds a minimum-robustness decomposition by solving a linear program in the Pauli transfer matrix representation~\cite{Bennink2017,Hakkaku2021}.
A shot samples one Clifford channel at each decomposed location with probability $|\alpha_k|/\mathcal R_{\mathrm{ch}}$ and records the corresponding sign.
The logical failure probability is estimated from the sign-weighted sample mean, multiplied by the product of the robustness factors over all sampled locations.
The estimator is unbiased, while its variance grows with the squared accumulated robustness.
Thus the cost of the method is controlled by the amount of non-Cliffordness, not directly by the number of qubits beyond the underlying stabilizer simulation.

Because this sampler decomposes the channel itself, it can represent non-unitary processes such as amplitude damping, as well as coherent channels.
Its flexibility can come at a higher robustness cost, especially when a coherent unitary contains off-diagonal terms that are expensive to synthesize from physical Clifford channels.

The unitary near-Clifford sampler is specialized to coherent errors of the form $\channel(\rho)=U\rho U^\dagger$.
Instead of decomposing the induced channel, it decomposes the operator,
\begin{equation}
    \label{eq:nc-unitary}
    U = \sum_i c_i\, C_i,
    \qquad
    \mathcal R_{\mathrm{op}} = \sum_i |c_i|,
\end{equation}
where $C_i$ are Clifford unitaries and $c_i$ are complex coefficients.
Substituting this expression into $U\rho U^\dagger$ produces cross terms $C_i\rho C_j^\dagger$.
For $i\neq j$ these are not physical Clifford channels, but they can be sampled by evolving the ket and bra sides with different Clifford unitaries and carrying the complex phase as a shot weight.
This requires a phase-sensitive stabilizer simulator rather than a standard tableau sampler: \plaq{} uses the C-H, or CH-form, representation of Ref.~\cite{Bravyi2019}, as in the operator-level near-Clifford method of Ref.~\cite{LeBlond2025}, so that ket-bra overlaps and measurement weights retain their relative Clifford phases.
The accumulated robustness factor is then the product of $\mathcal R_{\mathrm{op},i}^2$ over all unitary locations.

This operator-level route can be substantially more efficient for coherent errors, because it samples the off-diagonal structure directly rather than reconstructing it from signed physical channels.
The tradeoff is that it treats unitary non-Clifford content exactly, while non-unitary channels in the same task must be represented by a sampler-compatible approximation, such as a Pauli-twirled channel.
The channel-level and unitary near-Clifford samplers are therefore complementary.

The external exact near-Clifford routes wrap the \textit{tsim}~\cite{haenel2026tsim} and \textit{Clifft}~\cite{chase2026clifft} backends.
Rather than sampling a quasiprobability decomposition, \textit{tsim} uses stabilizer-rank and ZX-calculus techniques, while \textit{Clifft} factors the state into Clifford and Pauli frames plus a dynamic active state vector.
Their cost is governed by the residual non-Clifford structure after the Clifford part of the circuit has been simplified, rather than by the total number of Clifford operations.

An alternative approach to near-Clifford simulations is presented in Ref.~\cite{miller2025markovian}.
Here, Markovian noise is decomposed via a Taylor expansion.
The resulting method is a strong simulator, which returns probabilities of different outcomes, and extending this method to sampling logical error rates is an open problem.
Another approach to simulating near-Clifford coherent errors involves mapping stabilizer measurements into fermionic linear optics, which has been demonstrated to be performant and scalable for surface codes~\cite{bravyi2018correcting}.
Generalising it to other codes with different connectivity is an open problem.

\subsubsection{Full-state sampling on CPU and GPU}
\label{sec:fullstate}

The full-state samplers are the reference methods.
They do not reduce the channel to a Pauli, XPauli, or near-Clifford representation; instead, they propagate the state under the Kraus channels specified by the error model.
Their memory and runtime costs are exponential in the number of levels included in the full-state model: a qubit-only state vector scales as $2^n$, while leakage or mode simulations can introduce qudit or oscillator factors such as $3^n$ or larger, so they are used for small circuits and for validating the faster samplers.

The CPU backend builds on Cirq~\cite{cirq2024}.
In state-vector mode, noise is sampled by quantum trajectories: for each shot, the simulator samples one Kraus branch at each noisy location and propagates the resulting pure state.
This gives an unbiased Monte Carlo estimate of the noisy evolution.
In density-matrix mode, the simulator propagates the mixed state directly, applying the full channel $\rho\mapsto\sum_k K_k\rho K_k^\dagger$ at every noisy location.
This removes trajectory sampling error for a fixed circuit, but increases the state dimension from a vector to a density matrix.

The GPU backend builds on NVIDIA cuQuantum~\cite{bayraktar2023cuquantum} and is optimized for FTQC circuits that typically comprise large numbers of mid-circuit measurements and resets.
It represents the pure-state trajectory as a tensor network: the many-body state is stored as tensors connected by shared indices, and gates are applied by contracting gate tensors into that network.
Noise is sampled by trajectory unraveling.
At a noisy location, the backend computes the reduced density matrix of the target qubits, evaluates the Kraus-branch probabilities $p_k=\Tr(K_k\rho_{\mathrm{red}}K_k^\dagger)$, samples one branch, and applies the normalized operator $K_k/\sqrt{p_k}$.
The method therefore keeps the full channel at the level of trajectory sampling, while using tensor contractions to manipulate the state more efficiently on GPU hardware.

These full-state methods are not intended to replace the scalable samplers in large threshold studies.
Their role is to provide exact or trajectory-exact reference logical error rates against which the reduced representations can be compared.

\subsection{Decoders}
\label{sec:decoding}

Sampling produces measurement records and logical outcomes.
Decoding receives these records and returns a predicted logical value, which then enables computing the logical error rate.
\plaq{} deliberately separates this decoding problem from the sampling problem.
A sampler consumes the noisy circuit representation needed to generate records.
A decoder is initialized with a DEM, a classical stochastic model whose error mechanisms carry probabilities together with detector and logical signatures~\cite{gidney2021stim}.
A DEM can be constructed using the Pauli-twirled version of the noisy circuit.
The noisy circuit whose Pauli-twirled approximation is used for constructing the DEM could be the same as the sampled circuit, or it could be different to account for imperfect characterization of the hardware.
Alternatively, the decoder can be initialized using a DEM that is experimentally reconstructed using syndrome data on the hardware~\cite{blumekohout2025estimatingdem,takou2026logicaldem}.

At present, \plaq{} includes wrappers for the decoders PyMatching~\cite{higgott2022pymatching}, Fusion-Blossom~\cite{wu2023fusionblossom}, Chromobius~\cite{gidney2023chromobius}, Concat-MWPM~\cite{lee2025colorscaling}, BP-OSD~\cite{roffe2020decoding}, BP-LSD~\cite{hillmann2024localized}, Relay-BP~\cite{mueller2025relaybp}, belief matching~\cite{higgott_fragile_2022}, and Tesseract~\cite{beni2025tesseract}, each of which is initialized from a DEM.
For these decoders, the default decoder model is obtained by applying a location-wise Pauli twirl to the local channels of the physical error model~\cite{geller2013efficient,katabarwa2017dynamical}.
Equivalently, this retains the diagonal Pauli-transfer-matrix information defined in \cref{eq:ptm} and propagates the resulting stochastic Pauli model through the noiseless detector circuit.
This produces detector and logical signatures suitable for the decoder.

The decoder-side Pauli twirl is not necessarily the representation sampled by the circuit sampler.
For example, the XPauli sampler may track leaked levels, or a near-Clifford sampler may track coherent interference, while the decoder is still initialized from a Pauli-twirled DEM.
This is the current default decoder interface, not a statement that Pauli twirling is optimal for coherent, leakage, or history-dependent noise.
The separation leaves room for richer decoder models, such as correlated hypergraph models, fitted detector priors, leakage-aware state variables, trajectory-conditioned decoder models, or decoders tailored to coherent noise.

The output of the sampling-decoding workflow is a logical failure probability $P_{\mathrm{fail}}(\chi;\bm{\eta})$ for the chosen code instance, physical parameters, sampler, and decoder.
For ordinary Monte Carlo sampling, confidence intervals follow from binomial statistics of the observed logical failures.
For sign- or phase-weighted near-Clifford estimators, the uncertainty is computed from the standard error of the weighted mean.

\subsection{Loss-aware decoding}
\label{sec:lad}

Loss-aware decoding is a specialization of the decoder layer for delayed heralding.
A qubit may be lost before the circuit knows where the loss occurred.
Until a loss-detection unit heralds the event, the missing qubit can still affect later gates and syndrome measurements.
A decoder that treats only the final heralding event, and not the possible loss history that produced it, can assign the wrong detector weights and fail to correct otherwise correctable errors.

\plaq{} handles this situation with a loss-aware decoding pipeline.
The input is a hardware-model circuit that specifies where erasures may occur, how lost qubits interact with later operations, and where the loss-detection units are placed.
For each sampled trajectory, \plaq{} records possible loss locations and the observed heralding pattern.
It then constructs a trajectory-conditioned decoder-model circuit, converts it to a detector-error model, and decodes the syndrome against the matching graph appropriate to that trajectory.
This dynamically updates the decoder model using information that becomes available during the shot.
Using \plaq{}'s general and flexible loss-aware decoding pipeline, we are able to reproduce the results of Refs.~\cite{gu2025erasure,yu2026rydberg}.
An alternative approach to loss-aware decoding is presented in Ref.~\cite{baranes2026qubitlossdetection}.

\subsection{Analysis: estimating thresholds, surfaces, and suppression}
\label{sec:analysis}

The analysis layer considers individual estimates of $P_{\mathrm{fail}}(\chi;\bm{\eta})$ and returns the performance quantities needed for hardware design.
It parallelizes the sampling-decoding workflow across physical parameters, code instances, samplers, and decoders, and then aggregates the results into thresholds, threshold surfaces, and below-threshold suppression factors.

For a one-parameter threshold scan, \plaq{} evaluates the logical-failure curves $\{P_{\mathrm{fail}}(\chi_1;\eta),P_{\mathrm{fail}}(\chi_2;\eta),\ldots\}$ for a family of code instances.
One estimator locates the finite-distance crossing by interpolating these curves, using Akima interpolation in the current implementation, and then refining the sampling around the crossing~\cite{stephens2014faulttolerant}.
Statistical uncertainty is estimated by bootstrap resampling the Monte Carlo data at each simulated point, repeating the full crossing extraction, and reporting the spread of the resulting threshold estimates.
This uncertainty captures sampling noise; it does not include systematic error from finite code sizes, the interpolation window, or the assumed scaling behavior.

A complementary estimator is finite-size scaling.
Near threshold, the logical failure probability is expected to obey the scaling form
\begin{equation}
    P_{\mathrm{fail}}(\chi;\eta)
    \simeq
    F\!\left((\eta-\eta^\star)d^{1/\nu}\right),
\end{equation}
where $d$ is the code distance, $\nu$ is a critical exponent, and $F$ is a smooth scaling function~\cite{wang2003confinement,wang2010threshold,katzgraber2009color,bombin2012strong,watson2014logical}.
Fitting all sampled $(d,\eta)$ points returns the threshold $\eta^\star$ together with $\nu$.
The corresponding data collapse provides a check that the simulated distances and parameter window are close enough to the asymptotic scaling regime.
In this paper, the superconducting leakage example and the trapped-ion heating example use this finite-size-scaling (FSS) route.
We use bootstrapping over the distribution of individual logical error rate values (redrawing the error counts from a binomial distribution) and repeat FSS multiple times to obtain confidence intervals for threshold values~\cite{efron1979bootstrap}.

Below threshold, \plaq{} estimates the suppression factor $\Lambda$ of \cref{eq:suppression} when the code family is naturally indexed by distance.
The per-shot logical failure probability is converted to a per-round logical error rate, and the scaling form is fit by linear regression of $\log p_{L,d}^{\mathrm{round}}$ against $(d+1)/2$.
The fitted $\Lambda$ describes how rapidly logical failure is suppressed as the code grows and is therefore the quantity most directly connected to resource overhead.
For code families not naturally indexed by distance, the analysis compares the logical-failure curves directly rather than imposing this scaling form.
Threshold surfaces are mapped using the ray construction of \cref{sec:surface} and reported with the corresponding hardware results, where the axes have concrete physical meaning.

The framework is now complete.
A physical error model enters once, as a channel or open-system description.
\plaq{} compiles that model into the representation required by the chosen sampler and decoder, samples and decodes the resulting circuit, and analyzes the logical error rates as functions of physical parameters.
The next section demonstrates this path in four examples: comparison of sampler approximations against full-state references, leakage-aware threshold estimation for superconducting qubits, neutral-atom threshold surfaces under intermediate-state scattering, and heating dynamics in trapped-ion architectures.

\section{Results: from physical error models to thresholds}
\label{sec:usecases}

This section demonstrates the \plaq{} framework introduced above via examples related to designing FTQCs in the presence of realistic hardware imperfections.

We start by considering a simplified setting to introduce the framework with a single error model being simulated via different samplers.
XPauli and near-Clifford can accurately track leakage and coherence, respectively, yielding logical error rates that are close to full-state simulations; conventional Pauli-twirling techniques fall short in this simplified scenario because they can neither properly include coherence effects over multiple rounds nor errors introduced by leaked states.

Next, we consider realistic hardware-derived settings: leakage in superconducting qubits, intermediate-state scattering in neutral atoms, and heating dynamics in trapped ions, which can be modelled by describing the Lindbladian governing the system evolution.
We begin by considering few-qubit examples and show that XPauli results match those from full-state simulations, whereas Clifford-only results can differ.
We then switch to larger distance codes comparing XPauli with Clifford-only simulations and show that the threshold obtained with XPauli differs significantly from that obtained from a Clifford-only Pauli-twirled approximation, with logical error rates potentially differing by an order of magnitude or more.

\subsection{Logical error rates under simplified error models}
\label{sec:results_error_model}

This section provides an example of \plaq{}'s framework in action: defining an error model directly in terms of physical errors, e.g.\ via unitary matrices or Kraus channels, and simulating it with different samplers without any manual reformulation.
The goal of this subsection is to validate both our error-conversion pipeline, which translates physical errors into sampler-readable instructions, and the accuracy of \plaq{}'s XPauli and near-Clifford stabilizer simulations.

We consider the scenario of a memory experiment for the $X$ logical operator of a distance-three repetition code, performed over three stabilizer measurement rounds.
We choose this deliberately small code because the full-state reference must include all levels in the error model, not only the computational qubit states.
In the leakage case below, this means propagating three-level qudits, so the state-vector dimension grows as $3^n$ rather than with the $2^n$ scaling of a qubit-only reference.
The distance-three repetition code is therefore the smallest setting that allows comparison of the different samplers while keeping the full-state reference feasible.

We consider two separate sources of errors: \begin{enumerate*}[label=(\roman*), itemjoin={{, }}, itemjoin*={{, and }}]
    \item coherent single-qubit rotation at the beginning of each round
    \item leakage errors after each \textsc{CZ} gate
\end{enumerate*}.
Each error model is defined based on its physics representation, and then simulated using three different samplers, namely Stim, XPauli or near-Clifford, and full-state.
In all cases we use PyMatching~\cite{higgott2022pymatching} for decoding, which is initialized with a DEM constructed from the Pauli-twirled version of the noisy circuit.

The first error model is the coherent single-qubit over-rotation
\begin{equation}
    U = \exp\!\big[-i\,\theta\pi\,(X+Z)\big], \qquad \theta = 0.05,
    \label{eq:sampler-coherent}
\end{equation}
inserted as an error operation before every round and held fixed.
We compare the logical error rates of the full-state simulation, Stim, and the unitary near-Clifford sampler.
This sampler is chosen as it can track the coherence of the error throughout the evolution of the memory.
The simple Pauli twirl of \cref{eq:sampler-coherent} is the stochastic Pauli channel
\begin{equation}
    \channel_{\mathrm{PT}}(\rho) = (1 - 2p)\,\rho + p\,X\rho X + p\,Z\rho Z,
    \quad p \approx 0.0243,
    \label{eq:stim-coherent}
\end{equation}
with no $Y$ component and no off-diagonal weight, discarding the coherence entirely.
This is the error model that \plaq{} automatically constructs for Stim.
The unitary near-Clifford sampler instead expresses and samples $U$ over Clifford operators with complex coefficients (see \cref{eq:nc-unitary}), returning the three-term decomposition
\begin{equation}
    U = c_0\,I + c_1\,S + c_2\,V,
    \qquad V = \tfrac{1}{\sqrt{2}}\big(X + i\,I\big),
    \label{eq:nc-coherent}
\end{equation}
with $S=\sqrt{Z}$, $c_0 \approx 0.664$, $c_1 \approx 0.156\,(1-i)$, and $c_2 \approx -0.220\,i$.
This decomposition has operator robustness $\mathcal R_{\mathrm{op}} = \sum_i |c_i| \approx 1.10$, and the accumulated operator robustness factor of the circuit is $\mathcal R_{\mathrm{tot}}^{\mathrm{op}}\approx 5.98$ as defined in \Cref{sec:ncsamplers}.

The second error model we consider is the population transfer $\ket{11}\!\leftrightarrow\!\ket{02}$, i.e.\ leakage, that a \textsc{CZ} gate can drive on its target qubit.
We write this error as the two-Kraus channel on a pair of three-level qudits
\begin{equation}
\begin{split}
    K_1 &= \sqrt{p_t}\,\big(\ketbra{02}{11} + \ketbra{11}{02}\big),\qquad p_t=0.2 \\
    K_0 &= \sqrt{\identity - K_1^\dagger K_1},
\end{split}
    \label{eq:sampler-leakage}
\end{equation}
applied after every \textsc{CZ} gate.
We compare the logical error rates of the full-state simulation, Stim, and the XPauli sampler, due to its ability to efficiently track leakage to higher energy states and scattering back into computational states.
Restricted to the computational subspace, the Pauli twirl that Stim receives as input is
\begin{equation}
\begin{split}
    \channel_{\mathrm{PT}}(\rho) = (1 - 3q)\,\rho
    + q\big(&Z_1\rho Z_1 + Z_2\rho Z_2 \\
    &+ Z_1 Z_2\,\rho\, Z_1 Z_2\big),
\end{split}
    \label{eq:stim-leakage}
\end{equation}
with $q \approx 7\times 10^{-4}$, tracing out the possibility of escaping the computational subspace.
Instead, the generalized-Pauli twirl error model that XPauli samples from directly accounts for the transitions that transfer population into and out of $\ket{2}$, and moreover considers the Pauli errors on the control qubit when the target leaks.
The generalized Pauli-twirled transition table automatically constructed by \plaq{} is shown in \Cref{tab:xpauli_leakage}.
The table shows that, if both control and target qubits start from the computational subspace, then they stay in the same subspace with 95\% probability, potentially acquiring additional $Z$ errors.
However, with 5\% probability the target qubit leaks and, in that case, the control qubit can experience either an $X$ or a $Y$ error with a further 50\% probability.
The table also accounts for the probabilities that the target qubit stays leaked (90\%) or leaks down to the computational subspace (10\%), associated with the possible errors on the control qubit.

\begin{table}[t]
\centering
\caption[Generalized-Pauli twirl of the leakage channel]{Generalized-Pauli twirl of the channel \cref{eq:sampler-leakage}.
$p_{\mathrm{tr}}$ is the probability of the transition, while $p_{\mathrm{err}\mid\mathrm{tr}}$ is the conditional probability of the listed Pauli error given that transition.
For each initial level-label combination, the transition probabilities $p_{\mathrm{tr}}$ sum to 1.
Only nonzero transition and conditional-error entries are listed.
}
\label{tab:xpauli_leakage}
\begin{tabular}{@{}lccr@{}}
\toprule
Transition & $p_{\mathrm{tr}}$ & Pauli error & $p_{\mathrm{err}\mid\mathrm{tr}}$ \\
\midrule
$(\comp,\comp)\!\to\!(\comp,\comp)$ & 0.9500 & $IZ$ & 0.0007 \\
 &  & $ZI$ & 0.0007 \\
 &  & $ZZ$ & 0.0007 \\
\noalign{\vskip 0.45ex{\color{black!25}\hrule height 0.35pt}\vskip 0.45ex}
$(\comp,\comp)\!\to\!(\comp,2)$ & 0.0500 & $X{\cdot}$ & 0.0250 \\
 &  & $Y{\cdot}$ & 0.0250 \\
\midrule
$(\comp,2)\!\to\!(\comp,\comp)$ & 0.1000 & $X{\cdot}$ & 0.0500 \\
 &  & $Y{\cdot}$ & 0.0500 \\
\noalign{\vskip 0.45ex{\color{black!25}\hrule height 0.35pt}\vskip 0.45ex}
$(\comp,2)\!\to\!(\comp,2)$ & 0.9000 & $Z{\cdot}$ & 0.0028 \\
\bottomrule
\end{tabular}
\end{table}

In both experiments, we create a single noiseless circuit for the repetition \Code object, we define the \ErrorModel based on either \Cref{eq:sampler-coherent} or \Cref{eq:sampler-leakage}, and then we sample 5,000 shots with the full-state sampler and 200,000 shots with the others.
The reduced number of shots for the full-state sampler was chosen to allow a reasonably low simulation time, while still having acceptable error bars.
\plaq{} automatically converts the \ErrorModel representation depending on the different samplers, and extracts a decoding graph from the Pauli-twirled DEM.

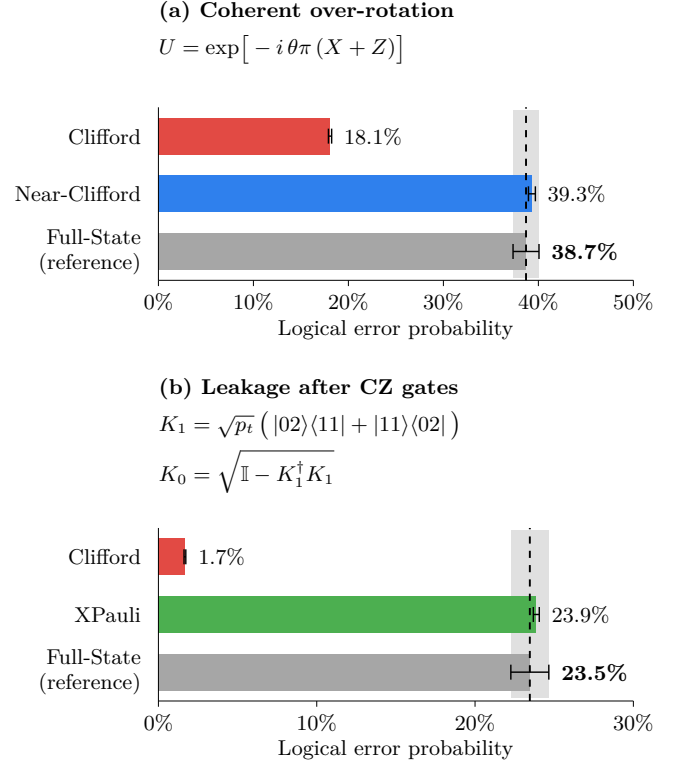
\begin{figure}[bt]
\centering
\resizebox{\columnwidth}{!}{%
\begin{tikzpicture}[font=\small]
\begin{scope}[x=14cm, y=0.85cm]
  \fill[fullstate!35] (0.373337,0.55) rectangle (0.400663,3.45);
  \fill[stim]         (0,2.68) rectangle (0.18068,3.32);
  \fill[nearclifford] (0,1.68) rectangle (0.393189,2.32);
  \fill[fullstate]    (0,0.68) rectangle (0.387,1.32);
  \draw[dashed,line width=0.8pt] (0.387,0.5) -- (0.387,3.5);
  \foreach \v/\e/\y in {0.18068/0.0016933/3, 0.393189/0.0036496/2, 0.387/0.013663/1}{%
    \draw[line width=0.6pt] (\v-\e,\y) -- (\v+\e,\y);
    \draw[line width=0.6pt] (\v-\e,\y-0.13) -- (\v-\e,\y+0.13);
    \draw[line width=0.6pt] (\v+\e,\y-0.13) -- (\v+\e,\y+0.13);
  }
  \node[right,xshift=2pt] at (0.18068+0.0016933,3) {18.1\%};
  \node[right,xshift=2pt] at (0.393189+0.0036496,2) {39.3\%};
  \node[right,xshift=2pt] at (0.387+0.013663,1) {\textbf{38.7\%}};
  \node[left,xshift=-4pt] at (0,3) {Clifford};
  \node[left,xshift=-4pt] at (0,2) {Near-Clifford};
  \node[left,xshift=-4pt,align=right] at (0,1) {Full-State\\(reference)};
  \draw (0,0.5) -- (0,3.5);
  \draw (0,0.5) -- (0.5,0.5);
  \foreach \x/\lab in {0/0, 0.1/10, 0.2/20, 0.3/30, 0.4/40, 0.5/50}{%
    \draw (\x,0.5) -- (\x,0.4);
    \node[below] at (\x,0.42) {\lab\%};
  }
  \node at (0.25,-0.35) {Logical error probability};
  \node[anchor=west, align=left, inner xsep=0pt, inner ysep=9pt]
    at (0,4.8) {\textbf{(a) Coherent over-rotation}\\[4pt]
      $U = \exp\!\big[-i\,\theta\pi\,(X+Z)\big]$};
\end{scope}

\begin{scope}[x=23.3333cm, y=0.85cm, yshift=-6.2cm]
  \fill[fullstate!35] (0.222601,0.55) rectangle (0.246599,3.45);
  \fill[stim]      (0,2.68) rectangle (0.016745,3.32);
  \fill[xpauli]    (0,1.68) rectangle (0.2387,2.32);
  \fill[fullstate] (0,0.68) rectangle (0.2346,1.32);
  \draw[dashed,line width=0.8pt] (0.2346,0.5) -- (0.2346,3.5);
  \foreach \v/\e/\y in {0.016745/0.00057/3, 0.2387/0.001875/2, 0.2346/0.011999/1}{%
    \draw[line width=0.6pt] (\v-\e,\y) -- (\v+\e,\y);
    \draw[line width=0.6pt] (\v-\e,\y-0.13) -- (\v-\e,\y+0.13);
    \draw[line width=0.6pt] (\v+\e,\y-0.13) -- (\v+\e,\y+0.13);
  }
  \node[right,xshift=2pt] at (0.016745+0.00057,3) {1.7\%};
  \node[right,xshift=2pt] at (0.2387+0.001875,2) {23.9\%};
  \node[right,xshift=2pt] at (0.2346+0.011999,1) {\textbf{23.5\%}};
  \node[left,xshift=-4pt] at (0,3) {Clifford};
  \node[left,xshift=-4pt] at (0,2) {XPauli};
  \node[left,xshift=-4pt,align=right] at (0,1) {Full-State\\(reference)};
  \draw (0,0.5) -- (0,3.5);
  \draw (0,0.5) -- (0.3,0.5);
  \foreach \x/\lab in {0/0, 0.1/10, 0.2/20, 0.3/30}{%
    \draw (\x,0.5) -- (\x,0.4);
    \node[below] at (\x,0.42) {\lab\%};
  }
  \node at (0.15,-0.35) {Logical error probability};
  \node[anchor=west, align=left, inner xsep=0pt, inner ysep=9pt]
    at (0,5.1) {\textbf{(b) Leakage after \textsc{CZ} gates}\\[4pt]
      $\begin{aligned}
        K_1 &= \sqrt{p_t}\,\big(\ketbra{02}{11} + \ketbra{11}{02}\big)\\
        K_0 &= \sqrt{\mathbb I - K_1^\dagger K_1}
      \end{aligned}$};
\end{scope}
\end{tikzpicture}}
\caption[Sampler comparison: accuracy and cost]{The XPauli and near-Clifford samplers reproduce the full-state logical error probabilities within confidence intervals, while results from \plaq{}'s Stim-based Clifford backend are significantly lower for these error models.
The comparison uses two non-Pauli imperfections evaluated at fixed strength on a distance-three repetition code, with 5,000 full-state shots and 200,000 shots for the other samplers.
Horizontal error bars and gray reference bands show two-sided 95\% intervals: Clopper--Pearson intervals for binomial samplers and weighted-estimator intervals for the near-Clifford sampler.
(a) Coherent over-rotation \cref{eq:sampler-coherent} at the start of each round;
(b) Leakage after \textsc{CZ}, as described by \cref{eq:sampler-leakage}.}
\label{fig:sampler_sweep}
\end{figure}

The results of the two comparisons are presented in \Cref{fig:sampler_sweep}.
The figure shows that the XPauli and near-Clifford samplers achieve logical error probabilities that are comparable with the full-state ones, while the Pauli-twirled circuit sampled via Stim provides a non-conservative error rate.
This example shows that a single error model, defined once in terms of its Kraus operators, suffices to run simulations with samplers at different levels of performance and accuracy.
We next consider a more realistic case of hardware-derived error models and observe the discrepancy between the different simulations there as well.

\subsection{Leakage in superconducting qubits}
\label{sec:leakage_superconducting}
In this example, we showcase how \plaq{} simulates leakage in superconducting qubits (transmons) starting directly from the Lindblad equation of motion that governs the physical evolution of the system.
Unlike conventional pipelines, \plaq{} does not require the user to supply a pre-computed Pauli-twirling approximation of the errors: the user specifies the physical evolution they want to model, and \plaq{} handles the conversion.
Starting from the interaction Hamiltonian and the Lindblad operators for dephasing, heating, and cooling, \plaq{} automatically extracts an error model that captures both Pauli errors and leakage, going beyond what a Pauli-twirling approximation can represent.
We use this model to compare the threshold obtained from the leakage-aware XPauli representation with the threshold obtained after reducing the same physical channel to a Clifford-only Pauli-twirled approximation and simulating that approximation with Stim.
We consider a model with a square pulse implementing a direct-coupling frequency-tuned \textsc{CZ} for simplicity~\cite{strauch2003quantum,dicarlo2009demonstration}, but the framework extends readily to tunable couplers with arbitrary pulse shapes.

A transmon qubit is a weakly anharmonic oscillator whose two lowest levels $\ket{0},\ket{1}$ form the computational subspace and whose third level $\ket{2}$ is the leading leakage channel.
A single three-level transmon is described by the Hamiltonian
\begin{equation}
    H_{\mathrm{trans}}(\omega,\alpha) = \omega\,a^\dagger a - \tfrac{\alpha}{2}\,a^\dagger a\,(a^\dagger a - 1),
    \label{eq:transmon}
\end{equation}
with the bosonic creation (annihilation) operator $a^\dagger$ ($a$), frequency $\omega$ and anharmonicity $\alpha$: the anharmonicity sets the detuning of the $\ket{1}\!\to\!\ket{2}$ transition from the $\ket{0}\!\to\!\ket{1}$ one, and it is this nonlinearity that both enables two-qubit gates and determines how easily the qubit leaks~\cite{koch2007charge}.

To implement the \textsc{CZ} gate, two transmons are coupled capacitively,
\begin{equation}
\begin{split}
    H = {}& H_{\mathrm{trans}}(\omega_1,\alpha_1)\otimes\identity
    + \identity\otimes H_{\mathrm{trans}}(\omega_2,\alpha_2) \\
    & - g\,(a_1^\dagger a_2 + a_1 a_2^\dagger),
\end{split}
    \label{eq:sc-twoqubit}
\end{equation}
with coupling strength $g$, and the \textsc{CZ} gate is realized by tuning the qubit frequencies so that $\ket{11}$ and $\ket{02}$ become resonant and letting the system evolve for a time $\tau_{\textsc{CZ}} = \pi/(\sqrt{2}\,|g|)$.
This imprints a $\pi$ phase on $\ket{11}$ while transiently populating $\ket{02}$, and it is exactly this excursion through $\ket{02}$ that can leave residual leaked population behind in $\ket{2}$ at the end of the gate.
Throughout this section the parameters of the Hamiltonian are fixed in dimensionless units at $\omega = 4.0$ for the untuned transmon, $\alpha_1 = \alpha_2 = 2.0$, and $g = 0.005$, so that $\tau_{\textsc{CZ}} \approx 444$.
Like the noise rates below, these values are illustrative rather than tuned to a specific device.

While the gate is being implemented, each transmon also couples to its environment through dephasing, heating, and cooling, described by the Lindblad operators
\begin{equation}
\begin{aligned}
    L_{\mathrm{deph}} &= \sqrt{\Gamma_{\mathrm{deph}}}\,a^\dagger a, &
    L_{\mathrm{heat}} &= \sqrt{\Gamma_{\mathrm{heat}}}\,a^\dagger, \\
    L_{\mathrm{cool}} &= \sqrt{\Gamma_{\mathrm{cool}}}\,a.
\end{aligned}
    \label{eq:sc-lindblad}
\end{equation}
Integrating the master equation of \cref{eq:lindblad} gives the imperfect implementation.
As described in \cref{sec:channels}, the error channel $\channel_{\bm{\eta}}$ is obtained by composing this imperfect implementation with the inverse of the target unitary gate $\mathcal G$, which in this case corresponds to the \textsc{CZ} gate.
When sampling with the XPauli sampler, \plaq{} converts the calculated error channel to its XPauli representation (\cref{sec:xpauli}) using a generalized Pauli twirl.
The resulting error model contains not only the leakage transitions from the computational subspace to $\ket{2}$ and back but also the Pauli errors, which would also be retrieved when performing Pauli twirling for pure stabilizer simulations.
For the Stim comparison, \plaq{} instead projects the same calculated channel onto the computational subspace and applies the ordinary Pauli twirl, yielding a Clifford-only stochastic Pauli model that Stim can sample.

\begin{figure}[t]
\centering
\begin{tikzpicture}
\pgfplotsset{
    sc physical cz axis/.style={
        width=0.95\columnwidth,
        height=0.62\columnwidth,
        xmin=0.00055,
        xmax=0.023,
        log basis x=10,
        xlabel={$\gamma_{\mathrm{heat}}=\Gamma_{\mathrm{heat}}\tau_{\mathrm{CZ}}$},
        ylabel={Logical error probability},
        xtick={0.000625,0.00125,0.0025,0.005,0.01,0.02},
        xticklabels={$6.25{\cdot}10^{-4}$,$1.25{\cdot}10^{-3}$,$2.5{\cdot}10^{-3}$,$5{\cdot}10^{-3}$,$10^{-2}$,$2{\cdot}10^{-2}$},
        grid=both,
        major grid style={black!14},
        minor grid style={black!8},
        tick label style={font=\tiny},
        label style={font=\small},
        title style={at={(0.02,1.02)}, anchor=south west, font=\small},
        legend style={
            at={(0.03,0.97)},
            anchor=north west,
            font=\sffamily\tiny,
            fill=white,
            fill opacity=0.86,
            draw=black!20,
            inner sep=1pt,
            cells={anchor=west}
        },
        legend cell align=left,
        unbounded coords=discard,
        filter discard warning=false,
    }
}
\begin{loglogaxis}[
    sc physical cz axis,
    name=scphysicalsmall,
    title={(a) $d=3$, full-state validation},
    ymin=7e-4,
    ymax=0.15,
    ytick={0.001,0.003,0.01,0.03,0.1},
    yticklabels={$10^{-3}$,$3{\cdot}10^{-3}$,$10^{-2}$,$3{\cdot}10^{-2}$,$10^{-1}$},
    xticklabel=\empty,
    xlabel={},
]
\addplot+[
    solid,
    color=stim,
    mark=triangle*,
    mark size=1.7pt,
    line width=0.75pt,
    mark options={fill=stim},
    restrict expr to domain={\thisrow{experiment_id}}{0:0},
    restrict expr to domain={\thisrow{sampler_id}}{1:1},
    restrict expr to domain={\thisrow{is_zero}}{0:0},
    forget plot,
    error bars/.cd,
        y dir=both,
        y explicit,
        error bar style={line width=0.65pt},
        error mark options={line width=0.65pt, mark size=2.3pt},
] table[x=gamma, y=plot_ler, y error plus=err_high, y error minus=err_low]
    {figures/generated/sc_physical_cz_heating_scan.dat};
\addplot+[
    solid,
    color=fullstate,
    mark=*,
    mark size=1.7pt,
    line width=0.75pt,
    mark options={fill=fullstate},
    restrict expr to domain={\thisrow{experiment_id}}{0:0},
    restrict expr to domain={\thisrow{sampler_id}}{2:2},
    restrict expr to domain={\thisrow{is_zero}}{0:0},
    forget plot,
    error bars/.cd,
        y dir=both,
        y explicit,
        error bar style={line width=0.65pt},
        error mark options={line width=0.65pt, mark size=2.3pt},
] table[x=gamma, y=plot_ler, y error plus=err_high, y error minus=err_low]
    {figures/generated/sc_physical_cz_heating_scan.dat};
\addplot+[
    solid,
    color=xpauli,
    mark=square*,
    mark size=1.7pt,
    line width=0.75pt,
    mark options={fill=xpauli},
    restrict expr to domain={\thisrow{experiment_id}}{0:0},
    restrict expr to domain={\thisrow{sampler_id}}{0:0},
    restrict expr to domain={\thisrow{is_zero}}{0:0},
    forget plot,
    error bars/.cd,
        y dir=both,
        y explicit,
        error bar style={line width=0.65pt},
        error mark options={line width=0.65pt, mark size=2.3pt},
] table[x=gamma, y=plot_ler, y error plus=err_high, y error minus=err_low]
    {figures/generated/sc_physical_cz_heating_scan.dat};
\addlegendimage{solid, color=stim, mark=triangle*, mark options={fill=stim}}
\addlegendentry{Clifford}
\addlegendimage{solid, color=xpauli, mark=square*, mark options={fill=xpauli}}
\addlegendentry{XPauli}
\addlegendimage{solid, color=fullstate, mark=*, mark options={fill=fullstate}}
\addlegendentry{Full state}
\node[anchor=south east, font=\tiny, text=black!55]
    at (rel axis cs:0.97,0.04) {95\% CI};
\end{loglogaxis}
\begin{loglogaxis}[
    sc physical cz axis,
    at={(scphysicalsmall.south west)},
    anchor=north west,
    yshift=-1.22cm,
    title={(b) $d=9$, larger code},
    ymin=1e-8,
    ymax=0.08,
    ytick={0.00000001,0.0000001,0.000001,0.00001,0.0001,0.001,0.01},
    yticklabels={$10^{-8}$,$10^{-7}$,$10^{-6}$,$10^{-5}$,$10^{-4}$,$10^{-3}$,$10^{-2}$},
    grid=major,
    minor tick style={draw=none},
]
\addplot+[
    solid,
    color=stim,
    mark=triangle*,
    mark size=1.7pt,
    line width=0.75pt,
    mark options={fill=stim},
    restrict expr to domain={\thisrow{experiment_id}}{1:1},
    restrict expr to domain={\thisrow{sampler_id}}{1:1},
    restrict expr to domain={\thisrow{is_zero}}{0:0},
    error bars/.cd,
        y dir=both,
        y explicit,
        error bar style={line width=0.65pt},
        error mark options={line width=0.65pt, mark size=2.3pt},
] table[x=gamma, y=plot_ler, y error plus=err_high, y error minus=err_low]
    {figures/generated/sc_physical_cz_heating_scan.dat};
\addplot+[
    solid,
    color=xpauli,
    mark=square*,
    mark size=1.7pt,
    line width=0.75pt,
    mark options={fill=xpauli},
    restrict expr to domain={\thisrow{experiment_id}}{1:1},
    restrict expr to domain={\thisrow{sampler_id}}{0:0},
    restrict expr to domain={\thisrow{is_zero}}{0:0},
    error bars/.cd,
        y dir=both,
        y explicit,
        error bar style={line width=0.65pt},
        error mark options={line width=0.65pt, mark size=2.3pt},
] table[x=gamma, y=plot_ler, y error plus=err_high, y error minus=err_low]
    {figures/generated/sc_physical_cz_heating_scan.dat};
\node[anchor=south east, font=\tiny, text=black!55]
    at (rel axis cs:0.97,0.04) {95\% CI};
\end{loglogaxis}
\end{tikzpicture}
\caption[Physical transmon-CZ heating validation]{Physical transmon-\textsc{CZ} error model in a heating-dominated regime.
The channel is derived from the Hamiltonian \cref{eq:sc-twoqubit} and the heating Lindblad operator in \cref{eq:sc-lindblad}, with dephasing and cooling set to zero and $\gamma_{\mathrm{heat}}=\Gamma_{\mathrm{heat}}\tau_{\textsc{cz}}$ swept logarithmically.
Both panels are shown on log-log axes.
Panel (a) uses a distance-three repetition-code $Z$-memory experiment over three rounds, sampled with $10^5$ shots for Stim and XPauli and $10^4$ shots for full-state.
Panel (b) uses a distance-nine repetition-code $Z$-memory experiment over nine rounds and no full-state leg.
Stim is sampled with $10^6$--$5\times10^8$ shots and XPauli with $10^6$--$5\times10^7$ shots, with the largest allocations assigned to lower resolved heating values where the confidence intervals are largest.
Colors and marker shapes encode the sampler.
Error bars show 95\% Clopper--Pearson confidence intervals; zero-count records are omitted from the plot.
XPauli tracks the full-state reference within the full-state sampling uncertainty in panel (a), while panel (b) shows that the Stim/XPauli separation persists for the larger code.
}
\label{fig:sc_physical_cz_heating_validation}
\end{figure}
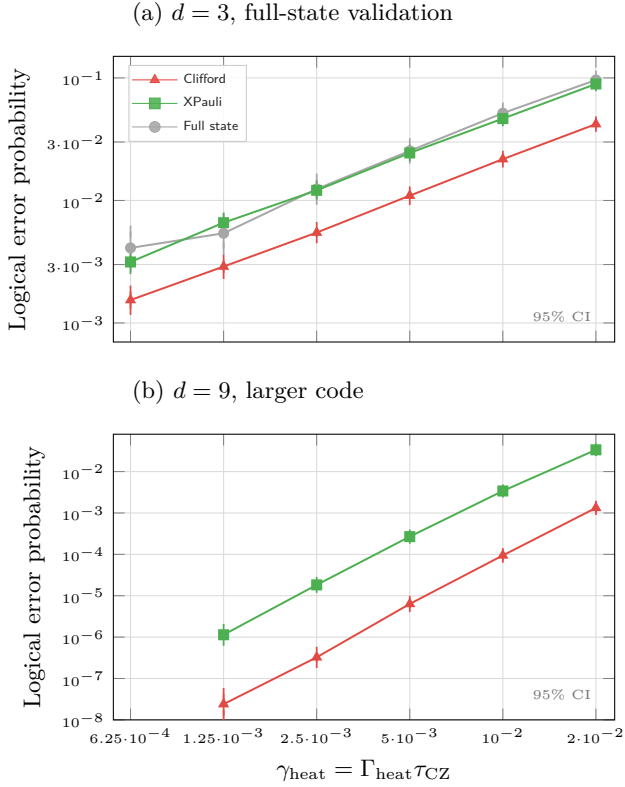

Before running the large-code threshold scan, we validate this physical qudit channel in a full-state-accessible setting and then repeat the same comparison at a larger code distance without the full-state sampler.
We set dephasing and cooling to zero, sweep the dimensionless heating strength $\gamma_{\mathrm{heat}}=\Gamma_{\mathrm{heat}}\tau_{\textsc{cz}}$, and run a distance-three repetition-code $Z$-memory experiment for three rounds with Stim, XPauli, and full-state sampling.
This heating-dominated scan isolates persistent leakage generated by the heating jump operator $a^\dagger$ and checks how closely the XPauli representation follows the exact Hilbert-space dynamics.
We also run a distance-nine repetition-code $Z$-memory experiment for nine rounds with Stim and XPauli only, which checks that the sampler separation is not an artifact of the small Hilbert-space benchmark.
The distance-nine comparison omits full-state sampling because retaining the transmon level structure already makes the exact qudit trajectory too expensive for this scan.
\Cref{fig:sc_physical_cz_heating_validation} shows that XPauli remains close to full-state in the distance-three panel while the Clifford-only Stim approximation gives smaller logical error probabilities, with the gap staying roughly a factor of two over the scanned range.
In the distance-nine panel, the resolved points where both samplers record failures place the XPauli logical error probability between $25$ and $56$ times above the Stim estimate, so the separation persists when the code is enlarged.

\begin{figure}[t]
\centering
\begin{tikzpicture}
\begin{semilogyaxis}[
    width=\columnwidth,
    height=0.78\columnwidth,
    xmin=0.0095,
    xmax=0.0225,
    ymin=7e-5,
    ymax=0.4,
    xlabel={Noise rate $\gamma$},
    ylabel={Logical error rate},
    scaled x ticks=false,
    xtick={0.010,0.014,0.018,0.022},
    xticklabel style={
        /pgf/number format/fixed,
        /pgf/number format/precision=3,
        font=\tiny
    },
    ytick={0.0001,0.001,0.01,0.1},
    yticklabels={$10^{-4}$,$10^{-3}$,$10^{-2}$,$10^{-1}$},
    grid=major,
    major grid style={black!14},
    tick label style={font=\tiny},
    label style={font=\small},
    legend style={
        at={(0.97,0.03)},
        anchor=south east,
        font=\tiny,
        fill=white,
        draw=black!20,
        inner sep=1pt,
        cells={anchor=west}
    },
    legend cell align=left,
    unbounded coords=discard,
    filter discard warning=false,
]
\addplot[draw=none, fill=xpauli, fill opacity=0.12, forget plot]
    coordinates {(0.0144513504,7e-5) (0.0152342655,7e-5) (0.0152342655,0.4) (0.0144513504,0.4)} -- cycle;
\addplot[draw=none, fill=stim, fill opacity=0.12, forget plot]
    coordinates {(0.0170697132,7e-5) (0.0183292810,7e-5) (0.0183292810,0.4) (0.0170697132,0.4)} -- cycle;
\addplot+[
    solid,
    color=xpauli!25!white,
    mark=square*,
    mark size=1.0pt,
    mark options={fill=xpauli!25!white},
    restrict expr to domain={\thisrow{sampler_id}}{0:0},
    restrict expr to domain={\thisrow{distance}}{5:5},
    forget plot,
    error bars/.cd,
        y dir=both,
        y explicit,
] table[x=gamma, y=ler, y error=sigma] {figures/generated/sc_leakage_threshold_ler.dat};
\addplot+[
    solid,
    color=xpauli!40!white,
    mark=square*,
    mark size=1.0pt,
    mark options={fill=xpauli!40!white},
    restrict expr to domain={\thisrow{sampler_id}}{0:0},
    restrict expr to domain={\thisrow{distance}}{7:7},
    forget plot,
    error bars/.cd,
        y dir=both,
        y explicit,
] table[x=gamma, y=ler, y error=sigma] {figures/generated/sc_leakage_threshold_ler.dat};
\addplot+[
    solid,
    color=xpauli!55!white,
    mark=square*,
    mark size=1.0pt,
    mark options={fill=xpauli!55!white},
    restrict expr to domain={\thisrow{sampler_id}}{0:0},
    restrict expr to domain={\thisrow{distance}}{9:9},
    forget plot,
    error bars/.cd,
        y dir=both,
        y explicit,
] table[x=gamma, y=ler, y error=sigma] {figures/generated/sc_leakage_threshold_ler.dat};
\addplot+[
    solid,
    color=xpauli!70!white,
    mark=square*,
    mark size=1.0pt,
    mark options={fill=xpauli!70!white},
    restrict expr to domain={\thisrow{sampler_id}}{0:0},
    restrict expr to domain={\thisrow{distance}}{11:11},
    forget plot,
    error bars/.cd,
        y dir=both,
        y explicit,
] table[x=gamma, y=ler, y error=sigma] {figures/generated/sc_leakage_threshold_ler.dat};
\addplot+[
    solid,
    color=xpauli,
    mark=square*,
    mark size=1.0pt,
    mark options={fill=xpauli},
    restrict expr to domain={\thisrow{sampler_id}}{0:0},
    restrict expr to domain={\thisrow{distance}}{13:13},
    forget plot,
    error bars/.cd,
        y dir=both,
        y explicit,
] table[x=gamma, y=ler, y error=sigma] {figures/generated/sc_leakage_threshold_ler.dat};
\addplot+[
    solid,
    color=xpauli!85!black,
    mark=square*,
    mark size=1.0pt,
    mark options={fill=xpauli!85!black},
    restrict expr to domain={\thisrow{sampler_id}}{0:0},
    restrict expr to domain={\thisrow{distance}}{15:15},
    forget plot,
    error bars/.cd,
        y dir=both,
        y explicit,
] table[x=gamma, y=ler, y error=sigma] {figures/generated/sc_leakage_threshold_ler.dat};
\addplot+[
    solid,
    color=xpauli!70!black,
    mark=square*,
    mark size=1.0pt,
    mark options={fill=xpauli!70!black},
    restrict expr to domain={\thisrow{sampler_id}}{0:0},
    restrict expr to domain={\thisrow{distance}}{17:17},
    forget plot,
    error bars/.cd,
        y dir=both,
        y explicit,
] table[x=gamma, y=ler, y error=sigma] {figures/generated/sc_leakage_threshold_ler.dat};
\addplot+[
    solid,
    color=xpauli!55!black,
    mark=square*,
    mark size=1.0pt,
    mark options={fill=xpauli!55!black},
    restrict expr to domain={\thisrow{sampler_id}}{0:0},
    restrict expr to domain={\thisrow{distance}}{19:19},
    forget plot,
    error bars/.cd,
        y dir=both,
        y explicit,
] table[x=gamma, y=ler, y error=sigma] {figures/generated/sc_leakage_threshold_ler.dat};
\addplot+[
    solid,
    color=stim!25!white,
    mark=triangle*,
    mark size=1.1pt,
    mark options={fill=stim!25!white},
    restrict expr to domain={\thisrow{sampler_id}}{1:1},
    restrict expr to domain={\thisrow{distance}}{5:5},
    forget plot,
    error bars/.cd,
        y dir=both,
        y explicit,
] table[x=gamma, y=ler, y error=sigma] {figures/generated/sc_leakage_threshold_ler.dat};
\addplot+[
    solid,
    color=stim!40!white,
    mark=triangle*,
    mark size=1.1pt,
    mark options={fill=stim!40!white},
    restrict expr to domain={\thisrow{sampler_id}}{1:1},
    restrict expr to domain={\thisrow{distance}}{7:7},
    forget plot,
    error bars/.cd,
        y dir=both,
        y explicit,
] table[x=gamma, y=ler, y error=sigma] {figures/generated/sc_leakage_threshold_ler.dat};
\addplot+[
    solid,
    color=stim!55!white,
    mark=triangle*,
    mark size=1.1pt,
    mark options={fill=stim!55!white},
    restrict expr to domain={\thisrow{sampler_id}}{1:1},
    restrict expr to domain={\thisrow{distance}}{9:9},
    forget plot,
    error bars/.cd,
        y dir=both,
        y explicit,
] table[x=gamma, y=ler, y error=sigma] {figures/generated/sc_leakage_threshold_ler.dat};
\addplot+[
    solid,
    color=stim!70!white,
    mark=triangle*,
    mark size=1.1pt,
    mark options={fill=stim!70!white},
    restrict expr to domain={\thisrow{sampler_id}}{1:1},
    restrict expr to domain={\thisrow{distance}}{11:11},
    forget plot,
    error bars/.cd,
        y dir=both,
        y explicit,
] table[x=gamma, y=ler, y error=sigma] {figures/generated/sc_leakage_threshold_ler.dat};
\addplot+[
    solid,
    color=stim,
    mark=triangle*,
    mark size=1.1pt,
    mark options={fill=stim},
    restrict expr to domain={\thisrow{sampler_id}}{1:1},
    restrict expr to domain={\thisrow{distance}}{13:13},
    forget plot,
    error bars/.cd,
        y dir=both,
        y explicit,
] table[x=gamma, y=ler, y error=sigma] {figures/generated/sc_leakage_threshold_ler.dat};
\addplot+[
    solid,
    color=stim!85!black,
    mark=triangle*,
    mark size=1.1pt,
    mark options={fill=stim!85!black},
    restrict expr to domain={\thisrow{sampler_id}}{1:1},
    restrict expr to domain={\thisrow{distance}}{15:15},
    forget plot,
    error bars/.cd,
        y dir=both,
        y explicit,
] table[x=gamma, y=ler, y error=sigma] {figures/generated/sc_leakage_threshold_ler.dat};
\addplot+[
    solid,
    color=stim!70!black,
    mark=triangle*,
    mark size=1.1pt,
    mark options={fill=stim!70!black},
    restrict expr to domain={\thisrow{sampler_id}}{1:1},
    restrict expr to domain={\thisrow{distance}}{17:17},
    forget plot,
    error bars/.cd,
        y dir=both,
        y explicit,
] table[x=gamma, y=ler, y error=sigma] {figures/generated/sc_leakage_threshold_ler.dat};
\addplot+[
    solid,
    color=stim!55!black,
    mark=triangle*,
    mark size=1.1pt,
    mark options={fill=stim!55!black},
    restrict expr to domain={\thisrow{sampler_id}}{1:1},
    restrict expr to domain={\thisrow{distance}}{19:19},
    forget plot,
    error bars/.cd,
        y dir=both,
        y explicit,
] table[x=gamma, y=ler, y error=sigma] {figures/generated/sc_leakage_threshold_ler.dat};
\addplot[xpauli, dashed, line width=1.1pt, forget plot]
    coordinates {(0.0148428079,7e-5) (0.0148428079,0.4)};
\addplot[stim, dashed, line width=1.1pt, forget plot]
    coordinates {(0.0176994971,7e-5) (0.0176994971,0.4)};
\addlegendimage{solid, color=stim, mark=triangle*, mark options={fill=stim}}
\addlegendentry{\textsf{Clifford}}
\addlegendimage{solid, color=xpauli, mark=square*, mark options={fill=xpauli}}
\addlegendentry{\textsf{XPauli}}
\addlegendimage{black, dashed, line width=1.1pt}
\addlegendentry{$\gamma_{\thr}$}
\end{semilogyaxis}
\end{tikzpicture}
\caption[Threshold shift under Pauli twirling]{Logical error probability of rotated planar codes of distance $d = 5-19$ under the superconducting error model, as a function of the dimensionless noise strength $\gamma$ controlling dephasing, heating, and cooling rates.
Solid green curves show the leakage-aware XPauli estimates, while solid red curves show the Clifford-only Pauli-twirled approximation simulated with Stim; lighter to darker shades encode increasing code distance.
The vertical dashed lines mark the finite-size-scaling threshold estimates, $\gamma_{\mathrm{th}} = 0.0148 \pm 0.0004$ for XPauli and $\gamma_{\mathrm{th}} = 0.0177 \pm 0.0006$ for Stim, with translucent bands showing the corresponding uncertainty intervals.
These estimates show that the Pauli-twirled approximation shifts the threshold upward.
At $d=19$, the Clifford-only approximation also changes the logical error probability by approximately an order of magnitude at the low-noise end of the scan.}
\label{fig:sc_leakage_threshold}
\end{figure}
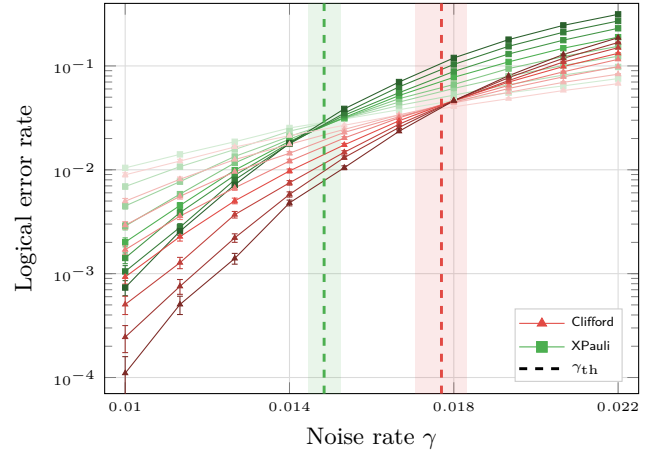

To calculate a combined error threshold for the dephasing, cooling, and heating noise sources, we scan a single dimensionless noise strength $\gamma$ that fixes the three dissipation rates relative to the gate duration, $\Gamma_{\mathrm{deph}} = \gamma/\tau_{\textsc{CZ}}$, $\Gamma_{\mathrm{heat}} = 0.02\,\gamma/\tau_{\textsc{CZ}}$, and $\Gamma_{\mathrm{cool}} = 0.1\,\gamma/\tau_{\textsc{CZ}}$, so that $\gamma$ measures the dissipation accumulated over one entangling gate.
Note that the numerical values here are chosen for illustrative purposes and values on actual hardware might differ from these.
For each value of $\gamma$ we build the gate error model, assemble the memory circuit of rotated planar codes of distance $d = 5-19$ run for $d$ rounds, and estimate the logical error probability of the logical $Z$ operator with both sampler representations over $2\times10^5$ shots per point.
Both representations are decoded with PyMatching~\cite{higgott2022pymatching}; XPauli readouts that end in leaked levels are resolved by random projection.
We concentrate the scan on the threshold region, sampling $\gamma$ at $10$ values linearly spaced in $[0.01, 0.022]$.
The logical error probabilities and thresholds of the system when scanning the noise parameter $\gamma$ are shown in \Cref{fig:sc_leakage_threshold}, with the XPauli estimates shown by solid green curves and the Stim estimates from the Clifford-only Pauli-twirled approximation shown by solid red curves.
The threshold is extracted using a finite-size-scaling (FSS) analysis of the logical error probability across all distances.
The threshold changes from $\gamma_{\mathrm{th}} = 0.0148 \pm 0.0004$ with XPauli to $\gamma_{\mathrm{th}} = 0.0177 \pm 0.0006$ with the Pauli-twirled approximation.
Thus, replacing leakage-aware sampling by the Pauli-twirled Clifford approximation shifts the estimated threshold upward by about 20\%.
For the largest simulated distance, $d=19$, the same approximation changes the logical error probability by approximately an order of magnitude at the low-noise end of the scan.

This example thus shows the \plaq{} framework in action: starting from the Hamiltonian of~\cref{eq:sc-twoqubit} and the Lindblad operators in~\cref{eq:sc-lindblad} alone, \plaq{} derives an XPauli-compatible error model, locates the threshold of the architecture, and enables quantifying how much the threshold changes when the same physical channel is reduced to a Clifford-only Pauli-twirled model.

\subsection{Threshold surfaces for neutral-atom imperfections}
\label{sec:neutral_atom_threshold_surface}

We next consider the example of neutral-atom Rydberg gates, where several different imperfections can act in tandem and their effect on FTQC performance needs to be studied carefully.
In this setting, we show that there is a significant distinction between results from Clifford-only and XPauli sampling.
We consider an illustrative model with square global pulses implementing the two-pulse \textsc{CZ} protocol of Ref.~\cite{levine2019parallel}, equal single-photon Rabi frequencies, and spontaneous emission decaying only to $\ket{1}$ for simplicity. 
For the same reason we scan the scattering rate $\gamma_{\mathrm{ISS}}$ directly, even though it is fixed by the atomic structure; the physically tunable knob is the intermediate-state detuning, which also controls the scattering-induced leakage and can equally well be scanned in \plaq{}. 
Moreover, the framework extends readily to shaped or time-optimal pulses, full radiative branching, other level schemes, and to atom loss with heralded erasure.

A neutral-atom qubit stores its computational states $\ket{0}$ and $\ket{1}$ in two hyperfine ground levels of a $^{87}$Rb atom, and the entangling gate is mediated by a highly excited Rydberg level $\ket{\mathrm{R}}$~\cite{saffman2010quantum,levine2019parallel,bluvstein2022quantum}.
The drive from $\ket{1}$ to $\ket{\mathrm{R}}$ is a two-photon process that passes through the intermediate level $\ket{\mathrm{6P}}$, and scattering from this intermediate level leads to leakage in this model.
In the interaction picture, the driven four-level atom is described by the Hamiltonian
\begin{equation}
\begin{split}
    H_{\mathrm{atom}}(\varphi) = {}& -\delta\,\ketbra{\mathrm{6P}}{\mathrm{6P}} - \Delta\,\ketbra{\mathrm{R}}{\mathrm{R}} \\
    & + \frac{\Omega}{2}\Big(e^{i\varphi}\ketbra{1}{\mathrm{6P}} + e^{i\varphi}\ketbra{\mathrm{6P}}{\mathrm{R}} + \mathrm{h.c.}\Big),
\end{split}
    \label{eq:na-atom}
\end{equation}
with Rabi frequency $\Omega$, laser phase $\varphi$, detuning $\delta$ of the drive from the intermediate level, and two-photon detuning $\Delta$ from the Rydberg level.
For $|\delta| \gg |\Omega|$ the intermediate level is populated only virtually, and the drive reduces to an effective two-level coupling between $\ket{1}$ and $\ket{\mathrm{R}}$ with Rabi frequency $\Omega_{\mathrm{eff}} = \Omega^2/(2\delta)$.
Single-qubit gates are driven in the same detuned two-photon fashion through a lower-lying intermediate level.

To implement the \textsc{CZ} gate, two atoms are placed within blockade range and driven simultaneously,
\begin{equation}
\begin{split}
    H = {}& H_{\mathrm{atom}}(\varphi)\otimes\identity + \identity\otimes H_{\mathrm{atom}}(\varphi) \\
    & + V_0\,\ketbra{\mathrm{RR}}{\mathrm{RR}},
\end{split}
    \label{eq:na-twoqubit}
\end{equation}
where the blockade energy $V_0$ shifts the doubly excited state $\ket{\mathrm{RR}}$ out of resonance, so that at most one of the two atoms can reach the Rydberg level.
The gate follows the two-pulse protocol of Ref.~\cite{levine2019parallel}: two global pulses of duration $\tau$ with a relative laser phase $\xi$, with $\Delta \approx 0.38\,\Omega_{\mathrm{eff}}$, $\xi \approx 3.90$, and $\tau \approx 4.29/\Omega_{\mathrm{eff}}$ chosen such that every computational state returns to itself and the accumulated phases satisfy $\varphi_{11} = 2\varphi_{01} + \pi$, which is a \textsc{CZ} gate up to single-qubit phases.
As for the transmon \textsc{CZ}, the gate transiently populates levels outside the computational subspace, here $\ket{\mathrm{6P}}$ and $\ket{\mathrm{R}}$, and it is this excursion that noise can turn into leakage.

While the pulses are applied, each atom also couples to its environment through dephasing of the computational levels and through spontaneous emission from the two excited levels, described by the Lindblad operators
\begin{equation}
\begin{aligned}
    L_{\mathrm{deph}} &= \sqrt{\gamma_{\mathrm{deph}}}\,\big(\ketbra{0}{0} - \ketbra{1}{1}\big), \\
    L_{\mathrm{ISS}} &= \sqrt{\gamma_{\mathrm{ISS}}}\,\ketbra{1}{\mathrm{6P}}, \\
    L_{\mathrm{R}} &= \sqrt{\gamma_{\mathrm{R}}}\,\ketbra{\mathrm{6P}}{\mathrm{R}},
\end{aligned}
    \label{eq:na-lindblad}
\end{equation}
where $\gamma_{\mathrm{deph}}$ is the dephasing rate and $\gamma_{\mathrm{ISS}}$ and $\gamma_{\mathrm{R}}$ are the inverse lifetimes of $\ket{\mathrm{6P}}$ and $\ket{\mathrm{R}}$.
Dephasing scrambles the relative phase of $\ket{0}$ and $\ket{1}$ and produces Pauli errors inside the computational subspace; intermediate-state scattering instead interrupts the coherent two-photon excursion, so that population which would have returned to the computational subspace ends the pulse sequence stranded in $\ket{\mathrm{6P}}$ or $\ket{\mathrm{R}}$.

We consider a rotated-planar surface-code memory experiment for the logical $Z$ operator, with code distance $d$ and $d$ stabilizer-measurement rounds.
As in the superconducting example, the gate error channels are computed from the driven dynamics of the platform: \plaq{} integrates the master equation \cref{eq:lindblad} for each gate pulse, with the Hamiltonian of \cref{eq:na-twoqubit} and the Lindblad operators of \cref{eq:na-lindblad}, and obtains the Kraus operators of the resulting map.
The parameters of the Hamiltonians are held fixed throughout the threshold surface scan.
Specifically, the \textsc{CZ} drive uses $\Omega = 2.7382$, $\delta = 300$, and $V_0 = 200$ in dimensionless units.
The three-dimensional parameter space in which we compute the threshold surface, is spanned by the following three scan parameters: \begin{enumerate*}[label=(\roman*), itemjoin={{; }}, itemjoin*={{; and }}]
    \item the depolarization probability $p_{\mathrm{depol}}$, which sets a qubit-level depolarizing channel applied before each stabilizer round
    \item the dephasing rate $\gamma_{\mathrm{deph}}$ of \cref{eq:na-lindblad}, which acts during the pulses of both the single-qubit gates and the entangling \textsc{CZ} gates
    \item the scattering rate $\gamma_{\mathrm{ISS}}$ of \cref{eq:na-lindblad}, which sets the decay of the intermediate $\ket{\mathrm{6P}}$ state through which the Rydberg gate passes, with the Rydberg-state decay tied to it at $\gamma_{\mathrm{R}} = 10^{-3}\,\gamma_{\mathrm{ISS}}$; this scattering is the dominant source of leakage out of the computational subspace
\end{enumerate*}.

The Pauli-twirling calculation projects each channel into the computational subspace and applies the ordinary Pauli twirl, yielding a Clifford-only stochastic Pauli model.
The XPauli calculation instead applies a generalized Pauli twirl, retaining the leakage transitions associated with the intermediate state.
Every scan point is a memory experiment for the logical $Z$ operator of rotated planar codes of distance $d = 9$--$15$ run for $d$ rounds, sampled with $2\times10^5$ shots and decoded with PyMatching~\cite{higgott2022pymatching}.

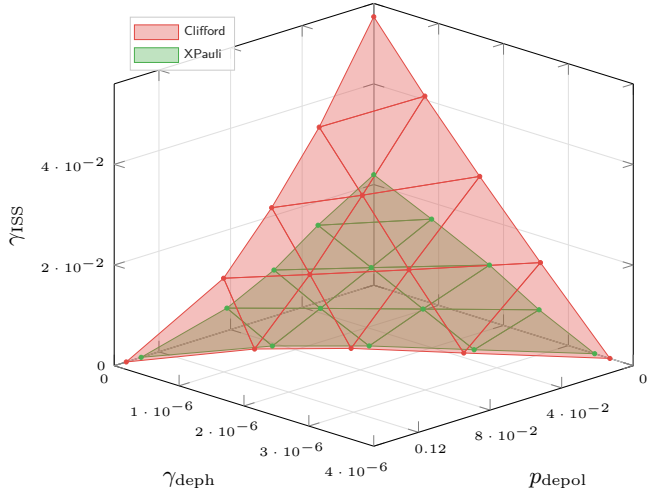
\begin{figure}
\centering
\resizebox{\columnwidth}{!}{%
\begin{tikzpicture}
\begin{axis}[
    width=\columnwidth,
    height=0.88\columnwidth,
    view={135}{22},
    xmin=0,
    xmax=0.145,
    ymin=0,
    ymax=4.0e-6,
    zmin=0,
    zmax=0.056,
    xlabel={$p_{\mathrm{depol}}$},
    ylabel={$\gamma_{\mathrm{deph}}$},
    zlabel={$\gamma_{\mathrm{ISS}}$},
    scaled ticks=false,
    xtick={0,0.04,0.08,0.12},
    ytick={0,1e-6,2e-6,3e-6,4e-6},
    ztick={0,0.02,0.04},
    tick label style={font=\tiny},
    label style={font=\small},
    grid=major,
    major grid style={black!12},
    legend style={
        at={(0.03,0.97)},
        anchor=north west,
        font=\tiny,
        fill=white,
        fill opacity=0.86,
        draw=black!20,
        inner sep=1.4pt,
        cells={anchor=west}
    },
    legend cell align=left,
]
\addplot3[
    patch,
    patch type=triangle,
    patch table={figures/generated/neutral_atom_threshold_surface/surface_patches_xpauli.dat},
    shader=flat,
    fill=xpauli,
    fill opacity=0.35,
    draw=xpauli,
    line width=0.25pt,
    forget plot,
] table[
    x=x_depol,
    y=y_deph,
    z=z_iss,
] {figures/generated/neutral_atom_threshold_surface/surface_vertices_xpauli.dat};
\addplot3[
    only marks,
    mark=*,
    mark size=0.8pt,
    color=xpauli,
    forget plot,
] table[
    x=x_depol,
    y=y_deph,
    z=z_iss,
] {figures/generated/neutral_atom_threshold_surface/surface_points_xpauli.dat};
\addplot3[
    patch,
    patch type=triangle,
    patch table={figures/generated/neutral_atom_threshold_surface/surface_patches_stim.dat},
    shader=flat,
    fill=stim,
    fill opacity=0.35,
    draw=stim,
    line width=0.25pt,
    forget plot,
] table[
    x=x_depol,
    y=y_deph,
    z=z_iss,
] {figures/generated/neutral_atom_threshold_surface/surface_vertices_stim.dat};
\addplot3[
    only marks,
    mark=*,
    mark size=0.8pt,
    color=stim,
    forget plot,
] table[
    x=x_depol,
    y=y_deph,
    z=z_iss,
] {figures/generated/neutral_atom_threshold_surface/surface_points_stim.dat};
\addlegendimage{area legend, fill=stim, draw=stim, fill opacity=0.35}
\addlegendentry{\textsf{Clifford}}
\addlegendimage{area legend, fill=xpauli, draw=xpauli, fill opacity=0.35}
\addlegendentry{\textsf{XPauli}}
\end{axis}
\end{tikzpicture}%
}
\caption[Neutral-atom threshold surfaces]{Threshold surfaces for the neutral-atom error model as a function of depolarizing error probability $p_{\mathrm{depol}}$, dephasing rate $\gamma_{\mathrm{deph}}$, and intermediate-state scattering rate $\gamma_{\mathrm{ISS}}$.
Markers show the per-ray threshold estimates over 15 barycentric scan directions; each surface interpolates the estimates of one sampler representation.
The red surface shows the Clifford-only Pauli-twirled approximation simulated with Stim; the green surface shows the leakage-aware XPauli calculation obtained from the generalized Pauli twirl.
Parameter combinations below a surface are below threshold for that representation.
The surfaces nearly coincide on the depolarization and dephasing axes and separate as the scattering weight grows, reaching a factor of $2.6$ on the $\gamma_{\mathrm{ISS}}$ axis; the volume between them is classified as correctable by the Clifford-only approximation but not by the leakage-aware XPauli calculation.}
\label{fig:neutral_atom_threshold_surface}
\end{figure}

\Cref{fig:neutral_atom_threshold_surface} shows that leakage sets the gap between the two threshold surfaces.
The surfaces nearly coincide on the depolarization and dephasing axes, but separate as scattering grows: on the scattering axis, the Clifford-only Pauli twirl drops leakage and overestimates the tolerable scattering rate by a factor of $2.6$.
Even in the $\gamma_{\mathrm{ISS}} = 0$ plane the two surfaces do not coincide exactly, because the gate channels retain a residual leakage component without any scattering: due to the finite blockade energy and detuning, the square pulses do not return the full population to the computational basis states, leaving residual leakage into $\ket{\mathrm{6P}}$ and $\ket{\mathrm{R}}$.
The Clifford-only projection discards this residual weight while the XPauli calculation retains it, which shifts the measured thresholds of the two representations apart by $7$--$11\,\%$ on these two axes.
The disagreement region is therefore concentrated at scattering-dominated parameter combinations, where the Clifford-only approximation classifies points as below threshold that XPauli classifies as above threshold.
The surface also exposes the interaction of the three noise sources: along the scattering-heavy interior direction with barycentric weights $(\tfrac14, \tfrac14, \tfrac12)$, the measured threshold lies at $0.86$ (Stim) and $0.93$ (XPauli) of the value obtained by linearly interpolating the single-axis thresholds, so the combined noise is mildly sub-additive.
Since the XPauli representation keeps the intermediate-state population dynamics that the Clifford-only projection removes, the XPauli surface is the more reliable design constraint for this neutral-atom regime.

\subsection{Trapped-ion heating}
\label{sec:trapped_ion_heating}

In the final example, we track how logical performance degrades in a trapped-ion architecture as the system heats up during the computation.
While the superconducting example considered an extra level involved in a gate, the trapped-ion example considers an extra `sector' that models the vibrational mode and whose state can change during the computation, thereby changing the error rates of later operations.
We simulate the heating dynamics using the Lindbladian equation of motion, which is then automatically converted to sector transitions in the XPauli extended state description.
Each sector is then associated with different depolarization rates, leading to logical errors that depend on the heating parameters of the system.
Such a scenario is not straightforwardly described within the fixed-probability stochastic Pauli noise that stabilizer simulations rely on, and thus showcases the flexibility of \plaq{} to describe complex interactions between the qubits and their environments.

\begin{figure}[t]
\centering
\resizebox{\columnwidth}{!}{%
\begin{tikzpicture}[
    font=\small,
    data/.style={circle, draw=iondata!55!ionline, fill=iondata, line width=0.7pt, minimum size=0.48cm, inner sep=0pt},
    xstab/.style={circle, draw=ionx!65!black, fill=ionx, text=white, line width=0.7pt, minimum size=0.48cm, inner sep=0pt},
    zstab/.style={circle, draw=ionz!65!black, fill=ionz, text=white, line width=0.7pt, minimum size=0.48cm, inner sep=0pt},
    level/.style={line width=1.05pt, line cap=round},
    heat/.style={-{Latex[length=1.7mm]}, line width=0.85pt, ionline},
    zoomline/.style={ionline!45, line width=0.7pt, dash pattern=on 2.6pt off 2.2pt},
]
    \begin{scope}[x=1.05cm, y=1.05cm, shift={(1.63,0)}]
        \fill[xpgreen] (0.75,2.15) rectangle (2.55,3.75);
        \fill[xpred]   (2.55,2.15) rectangle (4.35,3.75);
        \fill[xpred]   (0.75,0.55) rectangle (2.55,2.15);
        \fill[xpgreen] (2.55,0.55) rectangle (4.35,2.15);

        \fill[xpgreen] (2.55,3.75) arc[start angle=180, end angle=0, x radius=0.90, y radius=0.60] -- cycle;
        \draw[ionx, line width=0.9pt] (2.55,3.75) arc[start angle=180, end angle=0, x radius=0.90, y radius=0.60];
        \fill[xpred] (0.75,3.75) arc[start angle=90, end angle=270, x radius=0.60, y radius=0.80] -- cycle;
        \draw[ionz, line width=0.9pt] (0.75,3.75) arc[start angle=90, end angle=270, x radius=0.60, y radius=0.80];
        \fill[xpred] (4.35,2.15) arc[start angle=90, end angle=-90, x radius=0.60, y radius=0.80] -- cycle;
        \draw[ionz, line width=0.9pt] (4.35,2.15) arc[start angle=90, end angle=-90, x radius=0.60, y radius=0.80];
        \fill[xpgreen] (0.75,0.55) arc[start angle=180, end angle=360, x radius=0.90, y radius=0.60] -- cycle;
        \draw[ionx, line width=0.9pt] (0.75,0.55) arc[start angle=180, end angle=360, x radius=0.90, y radius=0.60];

        \draw[ionline!55, line width=0.9pt] (0.75,3.75) -- (4.35,3.75);
        \draw[ionline!55, line width=0.9pt] (0.75,2.15) -- (4.35,2.15);
        \draw[ionline!55, line width=0.9pt] (0.75,0.55) -- (4.35,0.55);
        \draw[ionline!55, line width=0.9pt] (0.75,3.75) -- (0.75,0.55);
        \draw[ionline!55, line width=0.9pt] (2.55,3.75) -- (2.55,0.55);
        \draw[ionline!55, line width=0.9pt] (4.35,3.75) -- (4.35,0.55);

        \node[data] (q0) at (0.75,3.75) {$0$};
        \node[data] (q1) at (2.55,3.75) {$1$};
        \node[data] (q2) at (4.35,3.75) {$2$};
        \node[data] (q3) at (0.75,2.15) {$3$};
        \node[data] (q4) at (2.55,2.15) {$4$};
        \node[data] (q5) at (4.35,2.15) {$5$};
        \node[data] (q6) at (0.75,0.55) {$6$};
        \node[data] (q7) at (2.55,0.55) {$7$};
        \node[data, line width=1.1pt] (q8) at (4.35,0.55) {$8$};

        \node[xstab] at (3.45,4.05) {$9$};
        \node[zstab] at (0.45,2.95) {$10$};
        \node[xstab] at (1.65,2.95) {$11$};
        \node[zstab] at (3.45,2.95) {$12$};
        \node[zstab] at (1.65,1.35) {$13$};
        \node[xstab] at (3.45,1.35) {$14$};
        \node[zstab] at (4.65,1.35) {$15$};
        \node[xstab] at (1.65,0.25) {$16$};

    \end{scope}

    \begin{scope}[x=1.05cm, y=1.05cm, shift={(0cm,-5.05cm)}]
        \coordinate (ionPanelNW) at (0.30,3.75);
        \coordinate (ionPanelNE) at (8.05,3.75);
        \filldraw[fill=ionpanel, draw=ionline, line width=0.9pt, rounded corners=7pt] (0.0,0.05) rectangle (8.35,3.75);

        \draw[ionline, level] (1.00,1.55) -- (2.45,1.55);
        \draw[ionline, level] (1.00,2.25) -- (2.45,2.25);
        \node[anchor=west] at (2.55,1.55) {$\ket{0}$};
        \node[anchor=west] at (2.55,2.25) {$\ket{1}$};
        \node[anchor=north] at (1.95,0.62) {electronic mode};

        \node[font=\Large] at (4.10,1.90) {$\otimes$};

        \draw[ionline, line width=1.1pt, smooth] plot[domain=-1.25:1.25, samples=61] ({6.05+\x},{0.70+1.696*\x*\x});
        \draw[linkblue, level]    (5.56,1.10) -- (6.54,1.10);
        \draw[linkblue!88, level] (5.34,1.55) -- (6.76,1.55);
        \draw[linkblue!76, level] (5.18,2.00) -- (6.93,2.00);
        \draw[linkblue!64, level] (5.03,2.45) -- (7.07,2.45);
        \draw[linkblue!52, level] (4.91,2.90) -- (7.19,2.90);
        \node[anchor=west, font=\scriptsize] at (6.68,1.10) {$\sigma=0$};
        \node[anchor=west, font=\scriptsize] at (6.90,1.55) {$\sigma=1$};
        \node[anchor=west, font=\scriptsize] at (7.07,2.00) {$\sigma=2$};
        \node[anchor=west, font=\scriptsize] at (7.21,2.45) {$\sigma=3$};
        \node[anchor=west, font=\scriptsize] at (7.33,2.90) {$\sigma=4$};
        \draw[heat] (6.05,1.16) -- (6.05,1.49);
        \draw[heat, opacity=0.55] (6.05,1.61) -- (6.05,1.94);
        \draw[heat, opacity=0.30] (6.05,2.06) -- (6.05,2.39);
        \draw[heat, opacity=0.16] (6.05,2.51) -- (6.05,2.84);
        \node[anchor=north] at (6.05,0.62) {vibrational mode};
    \end{scope}

    \draw[zoomline] (q8) -- (ionPanelNW);
    \draw[zoomline] (q8) -- (ionPanelNE);
\end{tikzpicture}%
}
\caption[Electronic and vibrational modes of a trapped ion]{
Each ion of the rotated planar code (left) carries an electronic mode that stores the qubit in the levels $\ket{0},\ket{1}$, tensored with a vibrational mode whose phonon-number levels are the sectors $\sigma=0,1,\dots$ (right).
Heating drives transitions upward through the vibrational ladder, which \plaq{} propagates as sector transitions while the electronic qubit evolves independently.
Errors on the electronic qubit can depend on the vibrational sector.
}
\label{fig:ion_modes}
\end{figure}
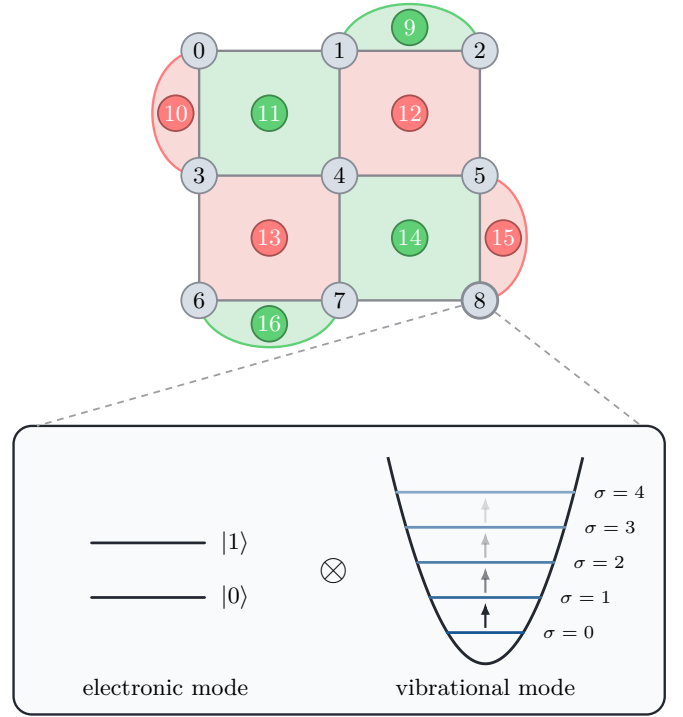

Trapped-ion qubits store information in long-lived electronic states, but their gates rely on shared motional, i.e., vibrational modes whose phonon occupation drifts upward as the ions exchange energy with the environment.
This \emph{heating} is not a Pauli error but is better modelled as a slow change of the environment that raises the error rate of subsequent gates.
\plaq{} captures it with a two-step approach: each qubit is attached to a classical sector label representing the state of the vibrational mode that evolves along with the computational state, while sector-dependent Pauli errors lead to errors affecting the computational state with a probability that depends on the local temperature of the mode.
In particular, we identify each sector $\sigma=0,1,2,\dots$ with the phonon-number state $\ket{\sigma}$ of the motional mode, so that the electronic degree of freedom carrying the qubit is decoupled from the vibrational degree of freedom carried by the sector (\Cref{fig:ion_modes}).
To define temperature-dependent Pauli errors, we define an XPauli error channel modelling depolarizing errors with strengths that increase with the sector label.

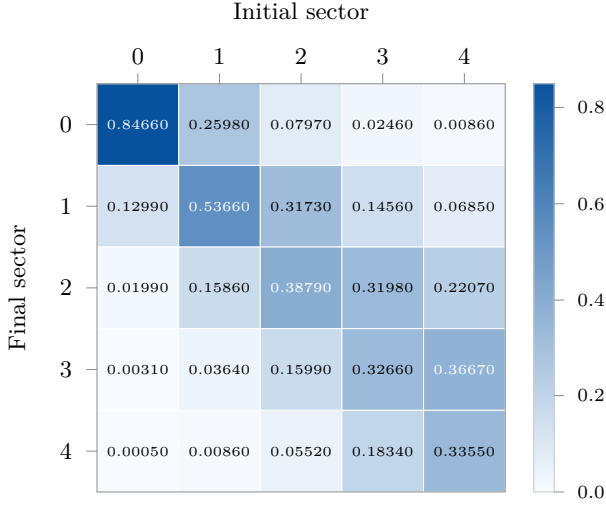
\begin{figure}[t]
\centering
\begingroup
\newcommand{\heatlabelcolor}[1]{\ifnum#1>42 white\else black\fi}
\newcommand{\heatcell}[4]{%
    \fill[draw=white, line width=0.25pt, fill=heathigh!#4!heatlow] (#1,-#2) rectangle ++(1,-1);
    \node[text=\heatlabelcolor{#4}, font=\fontsize{5.2}{6}\selectfont] at ({#1 + 0.5},{-#2 - 0.5}) {#3};
}
\begin{tikzpicture}[x=1.08cm, y=1.08cm]
    \heatcell{0}{0}{0.84660}{100}
    \heatcell{1}{0}{0.25980}{31}
    \heatcell{2}{0}{0.07970}{9}
    \heatcell{3}{0}{0.02460}{3}
    \heatcell{4}{0}{0.00860}{1}
    \heatcell{0}{1}{0.12990}{15}
    \heatcell{1}{1}{0.53660}{63}
    \heatcell{2}{1}{0.31730}{37}
    \heatcell{3}{1}{0.14560}{17}
    \heatcell{4}{1}{0.06850}{8}
    \heatcell{0}{2}{0.01990}{2}
    \heatcell{1}{2}{0.15860}{19}
    \heatcell{2}{2}{0.38790}{46}
    \heatcell{3}{2}{0.31980}{38}
    \heatcell{4}{2}{0.22070}{26}
    \heatcell{0}{3}{0.00310}{0}
    \heatcell{1}{3}{0.03640}{4}
    \heatcell{2}{3}{0.15990}{19}
    \heatcell{3}{3}{0.32660}{38}
    \heatcell{4}{3}{0.36670}{43}
    \heatcell{0}{4}{0.00050}{0}
    \heatcell{1}{4}{0.00860}{1}
    \heatcell{2}{4}{0.05520}{6}
    \heatcell{3}{4}{0.18340}{22}
    \heatcell{4}{4}{0.33550}{39}

    \draw[black!45, line width=0.35pt] (0,0) rectangle (5,-5);
    \foreach \sector in {0,...,4} {
        \draw[black!45, line width=0.35pt] ({\sector + 0.5},0) -- ({\sector + 0.5},0.13);
        \draw[black!45, line width=0.35pt] (0,{-\sector - 0.5}) -- (-0.13,{-\sector - 0.5});
        \node[font=\small] at ({\sector + 0.5},0.34) {$\sector$};
        \node[font=\small] at (-0.38,{-\sector - 0.5}) {$\sector$};
    }
    \node[font=\small] at (2.5,0.9) {Initial sector};
    \node[font=\small, rotate=90] at (-0.98,-2.5) {Final sector};

    \shade[bottom color=heatlow, top color=heathigh, draw=black!20]
        (5.35,-5) rectangle (5.6,0);
    \foreach \value/\ypos in {0.0/-5.00,0.2/-3.82,0.4/-2.65,0.6/-1.47,0.8/-0.29} {
        \draw[black!45, line width=0.25pt] (5.6,\ypos) -- (5.72,\ypos);
        \node[font=\scriptsize, anchor=west] at (5.78,\ypos) {\value};
    }
\end{tikzpicture}
\endgroup
\caption[Sector transitions of the heating channel]{Sector-transition probabilities of the heating channel, computed by \plaq{} from the Lindbladian \cref{eq:ion-lindblad} with $\Gamma_h = 0.2$ and $n_{\mathrm{th}} = 1.0$.
Sectors $0$--$4$ represent the phonon-number states $\ket{0}$--$\ket{4}$ of the motional mode.
Each column gives the probability of ending in a given sector after one idling interval; the population drifts upward toward higher phonon numbers.}
\label{fig:ion_heating_channel}
\end{figure}

In this modelling, heating is the transport of population toward higher sectors driven by a thermal bath, with absorption at rate $\Gamma_h\,n_{\mathrm{th}}$ and emission at rate $\Gamma_h\,(n_{\mathrm{th}}+1)$, where $\Gamma_h$ is the thermalization rate and $n_{\mathrm{th}}$ the mean phonon occupation at equilibrium, which is set by the system temperature.
The motional density matrix evolves under the Lindbladian
\begin{equation}
\begin{split}
    \mathcal{L}[\rho] = \;& \Gamma_h\,n_{\mathrm{th}}\Big( a^\dagger \rho\, a - \tfrac12\{a a^\dagger, \rho\} \Big) \\
    + \;& \Gamma_h\,(n_{\mathrm{th}}+1)\Big( a \rho\, a^\dagger - \tfrac12\{a^\dagger a, \rho\} \Big).
\end{split}
    \label{eq:ion-lindblad}
\end{equation}
\plaq{} obtains the heating channel by propagating \cref{eq:ion-lindblad} over an idling interval $\Delta t$, $\rho \mapsto e^{\mathcal{L}\Delta t}[\rho]$, and reading off the Kraus operators of the resulting map.
Because the bath couples only to the motional mode, these Kraus operators take a simple closed form: indexed by the initial and final sectors $\sigma, \sigma'$, they are
\begin{equation}
\begin{split}
    K_{\sigma'\leftarrow \sigma} &= \sqrt{T_{\sigma'\leftarrow \sigma}}\;\ketbra{\sigma'}{\sigma}\otimes\identity_{\comp}, \\
    T_{\sigma'\leftarrow \sigma} &= \bra{\sigma'} e^{\mathcal{L}\Delta t}\!\ketbra{\sigma}{\sigma} \ket{\sigma'},
\end{split}
    \label{eq:heating-kraus}
\end{equation}
where $\ketbra{\sigma'}{\sigma}$ acts on the vibrational mode, $\identity_{\comp}$ is the identity on the two computational levels, and $T_{\sigma'\leftarrow \sigma}$ is the probability of the corresponding sector transition.
The heating channel is therefore a classical stochastic transport on the phonon sectors that leaves the electronic qubit untouched.
The full sector-transition matrix $T$ is shown in \Cref{fig:ion_heating_channel} for $\Gamma_h = 0.2$, $n_{\mathrm{th}} = 1.0$, and five sectors: every column redistributes population from a given starting sector toward its neighbours, with a net upward drift that drives the ions toward thermal equilibrium.

The sector label only leads to an error once it is coupled to the computational state.
We model this with a sector-dependent depolarizing channel applied at the beginning of every stabilizer measurement round and at every two-qubit gate.
In our simplified error modelling, the depolarization probability $p_\mathrm{depol}$ is linearly dependent on the current sector's phonon number $n$ according to the relation
\begin{equation}
    p_\mathrm{depol}(n) = p_0+\kappa\left(2 n+1\right).
    \label{eq:heating_noise}
\end{equation}
We take $p_0 = 10^{-4}$ and $\kappa = 5\times10^{-3}$, so that the depolarizing probability grows from about $0.5\%$ in sector $\sigma=0$ to $4.5\%$ in sector $\sigma=4$.
This profile is a modelling relation based on Ref.~\cite{sutherland2022motional} and is used here to showcase \plaq{}'s ability to treat temperature-dependent depolarization errors; nevertheless, \plaq{} can accept any other sector-to-rate assignment equally well.
Heating and the sector-dependent depolarizing channels are combined into a single error model: the heating channel acts during idling before each round, while the depolarizing channel is applied at the beginning of each round and on the two-qubit gates, with a rate $p_\mathrm{depol}$ set by each qubit's current sector.

\begin{figure}[t]
\centering
\begin{tikzpicture}
\begin{semilogyaxis}[
    width=\columnwidth,
    height=0.78\columnwidth,
    xmin=0.00815,
    xmax=0.01285,
    ymin=0.018,
    ymax=0.58,
    xlabel={Average depolarizing rate $\bar p(\Gamma_h,d)$},
    ylabel={Logical error rate},
    scaled x ticks=false,
    xtick={0.0085,0.0095,0.0105,0.0115,0.0125},
    xticklabel style={
        /pgf/number format/fixed,
        /pgf/number format/precision=4,
        font=\tiny
    },
    ytick={0.03,0.1,0.3},
    yticklabels={$3{\times}10^{-2}$,$10^{-1}$,$3{\times}10^{-1}$},
    grid=both,
    major grid style={black!14},
    minor grid style={black!8},
    tick label style={font=\tiny},
    label style={font=\small},
    legend style={
        at={(0.97,0.03)},
        anchor=south east,
        font=\tiny,
        fill=white,
        fill opacity=0.86,
        draw=black!20,
        inner sep=1pt,
        cells={anchor=west}
    },
    legend cell align=left,
    unbounded coords=discard,
    filter discard warning=false,
]
\addplot[draw=none, fill=xpauli, fill opacity=0.12, forget plot]
    coordinates {(0.01079,0.018) (0.01095,0.018) (0.01095,0.58) (0.01079,0.58)} -- cycle;
\addplot+[
    solid,
    color=xpauli!35!white,
    mark=square*,
    mark options={fill=xpauli!35!white},
    restrict expr to domain={\thisrow{distance}}{9:9},
    error bars/.cd,
        y dir=both,
        y explicit,
] table[x=pbar, y=ler, y error=sigma] {figures/generated/ion_heating_xpauli_ler.dat};
\addlegendentry{$d=9$}
\addplot+[
    solid,
    color=xpauli!50!white,
    mark=square*,
    mark options={fill=xpauli!50!white},
    restrict expr to domain={\thisrow{distance}}{11:11},
    error bars/.cd,
        y dir=both,
        y explicit,
] table[x=pbar, y=ler, y error=sigma] {figures/generated/ion_heating_xpauli_ler.dat};
\addlegendentry{$d=11$}
\addplot+[
    solid,
    color=xpauli!70!white,
    mark=square*,
    mark options={fill=xpauli!70!white},
    restrict expr to domain={\thisrow{distance}}{13:13},
    error bars/.cd,
        y dir=both,
        y explicit,
] table[x=pbar, y=ler, y error=sigma] {figures/generated/ion_heating_xpauli_ler.dat};
\addlegendentry{$d=13$}
\addplot+[
    solid,
    color=xpauli,
    mark=square*,
    mark options={fill=xpauli},
    restrict expr to domain={\thisrow{distance}}{15:15},
    error bars/.cd,
        y dir=both,
        y explicit,
] table[x=pbar, y=ler, y error=sigma] {figures/generated/ion_heating_xpauli_ler.dat};
\addlegendentry{$d=15$}
\addplot+[
    solid,
    color=xpauli!75!black,
    mark=square*,
    mark options={fill=xpauli!75!black},
    restrict expr to domain={\thisrow{distance}}{17:17},
    error bars/.cd,
        y dir=both,
        y explicit,
] table[x=pbar, y=ler, y error=sigma] {figures/generated/ion_heating_xpauli_ler.dat};
\addlegendentry{$d=17$}
\addplot+[
    solid,
    color=xpauli!55!black,
    mark=square*,
    mark options={fill=xpauli!55!black},
    restrict expr to domain={\thisrow{distance}}{19:19},
    error bars/.cd,
        y dir=both,
        y explicit,
] table[x=pbar, y=ler, y error=sigma] {figures/generated/ion_heating_xpauli_ler.dat};
\addlegendentry{$d=19$}
\addplot[xpauli, dashed, line width=1.2pt]
    coordinates {(0.01087,0.018) (0.01087,0.58)};
\addlegendentry{$p_{\thr}$}
\end{semilogyaxis}
\end{tikzpicture}
\caption[XPauli heating logical-error-rate scan]{Logical error rate of rotated planar-code memory experiments against the round-averaged depolarizing rate per noise location.
Green squares show the XPauli heating scan, with each point labelled by the analytic round-averaged rate $\bar p(\Gamma_h, d)$ of its non-stationary noise.
The vertical line marks the finite-size-scaling threshold $p_{\thr} = 0.01087 \pm 0.00008$ for the heating data.
The heating data continue to show suppression with growing distance across the sampled rate window, reflecting that no single stationary effective rate reproduces the non-stationary dynamics.}
\label{fig:ion_heating_ler_comparison}
\end{figure}
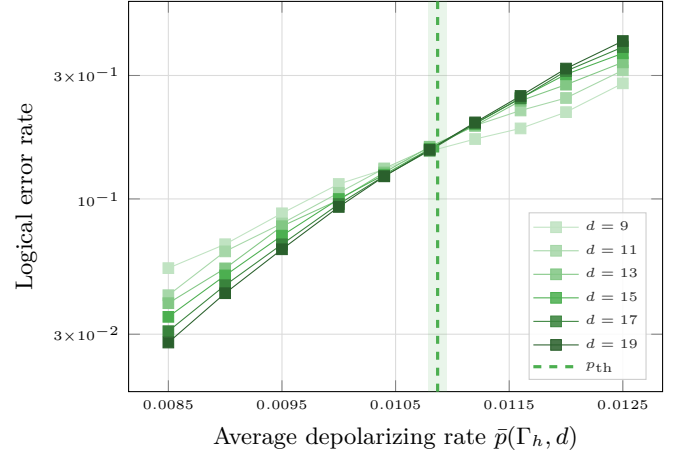

We then scan the thermalization rate $\Gamma_h$ over nine values $\Gamma_h \in [0.0229, 0.0429]$; for each value, we assemble memory circuits of rotated planar codes of distance $d = 9-23$ run for $d$ rounds, with every qubit starting in the motional ground state, $n_{\mathrm{th}} = 1$, and the phonon ladder truncated at five sectors.
We estimate the logical error probability of the logical $Z$ operator with the XPauli sampler over $2\times10^5$ shots per point, decoded with PyMatching~\cite{higgott2022pymatching}.
We use this scan to compare the sector-resolved heating model against the stationary Pauli approximation a Clifford-only simulation would normally use: XPauli propagates the phonon-sector history through each memory circuit, while a matching Stim control uses a fixed depolarizing rate throughout with identical average depolarizing probability.
The resulting threshold is presented in \cref{fig:ion_heating_ler_comparison}.

For concreteness, we have considered this simplified setting, which could potentially be simulated by considering Clifford circuits with depolarization rates increasing in subsequent rounds; the plot obtained shows the flexibility of the \plaq{} framework.
This flexibility could enable answering design questions such as how often it is optimal to cool ions or how motion impacts FTQC performance.

\section{Conclusions and outlook}
\label{sec:discussion}

Accurate and reliable modelling of hardware imperfections is fundamental to the design of architectures for FTQCs.
To this aim, \plaq{} is built on a single premise: that the noise of a quantum computer should be described once, as a CPTP channel derived from the device's physics, and then simulated under whatever approximation the question at hand can afford.
In this paper, we have discussed how the software realizes this premise: with a channel object that converts between six exact or approximate representations; a family of samplers that trade accuracy for scalability; and a decoding layer that is decoupled from sampling.

We have presented new samplers that enable simulations of hardware imperfections where both pure stabilizer-based and full-state methods usually cannot be used: the former can provide unreliable logical error rates that are often too optimistic, while the latter scale exponentially in the retained Hilbert-space dimension, including leakage levels or other qudit sectors when the physical model keeps them.
The XPauli sampler's efficiency rests on the assumption that all additional levels outside of the computational subspace do not exhibit coherence effects.
This often leads to an approximation of the hardware error model, but allows phenomena like leakage and shifting environmental conditions to be modelled with a high level of accuracy.
The near-Clifford samplers are instead exact, but pay a variance cost that grows with the squared robustness of the channel decomposition, so strongly non-Clifford noise may demand many shots.
These samplers are ideal for studying coherent errors and unitary error channels, whose superposition throughout the circuit cannot be easily accounted for using a simple stochastic approximation.

Spatially correlated crosstalk can be represented when it is supplied as a CPTP channel on the qudits involved at a circuit location.
Non-Markovian effects can also be simulated by embedding the relevant memory into the simulated state, for example as classical sectors or explicit environment levels.
However, arbitrary history-dependent quantum noise is outside the present circuit-level channel formalism.

We have shown \plaq{}'s capabilities for modelling real hardware error models by simulating leakage effects in superconducting qubits starting from the Lindbladian equation describing their evolution when implementing an entangling gate; by mapping neutral-atom threshold surfaces under intermediate-state scattering; by locating the threshold of a non-stationary heating model in trapped-ion architectures; and by comparing simulations of coherent errors and leakage under different approximations to demonstrate the accuracy of the different simulators.

\plaq{} thus has the potential to be an invaluable tool in the hands of quantum hardware designers, providing a direct path from an open-system model of a real device to the logical performance of a fault-tolerant architecture built from it, across every major hardware platform and computational paradigm, under realistic hardware imperfections.

\plaq{} addresses the challenge that hardware teams face: large-scale QEC studies often use stochastic Pauli approximations because they are compatible with fast stabilizer simulation, while real hardware noise can have richer structure.
In doing so, it helps hardware teams answer the practical questions of whether, for a given device, the noise is below threshold and by how much, and which physical imperfection most limits the available margin.
This informs design decisions that will enable hardware teams to reach fault tolerance faster.

\section*{Data availability}

The data supporting the findings of this study are available from the authors upon reasonable request.
The \plaq{} software is a commercial product of QC Design; information on access is available at \url{https://www.qc.design}.

\section*{Acknowledgements}
\plaq{} has benefited tremendously from feedback that we received from our customers and partners, and we are very grateful for it.
We thank Aggie Branczyk for help with the presentation of this manuscript.

\bibliographystyle{apsrev4-2}
\nocite{apsrev42Control}
\bibliography{references}

\end{document}